# Performing Creativity With Computational Tools

**Daniel F. S. Lopes**
University of Coimbra,
CISUC, DEI
dfl@dei.uc.pt

**Jéssica Araújo Parente**
University of Coimbra,
CISUC, DEI
jparente@dei.uc.pt

**Pedro Parreira e Silva**
University of Coimbra,
CISUC, DEI
pedros@dei.uc.pt

**Licínio Roque**
University of Coimbra,
CISUC, DEI
lir@dei.uc.pt

**Penousal Machado**
University of Coimbra,
CISUC, DEI
machado@dei.uc.pt

## Abstract

The introduction of new tools in people's workflow has always been promotive of new creative paths. This paper discusses the impact of using computational tools in the performance of creative tasks, especially focusing on graphic design. The study was driven by a grounded theory methodology, applied to a set of semi-structured interviews, made to twelve people working in the areas of graphic design, data science, computer art, music and data visualisation. Among other questions, the results suggest some scenarios in which it is or it is not worth investing in the development of new intelligent creativity-aiding tools.

## I. INTRODUCTION

Looking backwards at the history of humanity or simply making a retrospective into our daily creative practices, it is noticeable that the use of different tools often introduces new creative possibilities. Likewise, using different computational tools may result in similar results. This experiment intended to answer the hypothesis of "How creativity is impacted by computational tools, from non-computational tools to creativity support tools, co-creative tools and fully autonomous tools". It also matters to refer that this research was specially directed to the scope of design.

The first step to answer our hypothesis was to search about: (i) Creativity and the evolution of design tools (computational and non-computation); (ii) Creative systems outside the design field; and (iii) studies about how to enhance creativity.

Then, a set of semi-structured interviews was made with graphic designers and computer artists working in design, data science, computer art, music and data visualisation. Many were experienced in more than one of these areas. Lastly, the answers were analysed and discussed within the research group, leading to the results presented in this paper. Those turned out to be eclectic enough to respond, not only to our research question but also to several other questions. For instance, whether or not it is worth it to invest in the development of new computational tools for enhancing creativity — with special attention for co-creative and fully autonomous systems.

## II. BACKGROUND

### a. Creativity and the evolution of design tools

**Designing with non-computational tools**
Over time, graphic design suffered severe improvements. Since the beginning of design and artistic creation, there was always a need for doing something novel and creative. Different areas pinpoint very heterogeneous examples of what it is to be creative. That makes this concept a very difficult one to define with high consensuality (Veale et.al., 2019). Even though, novelty is most of the time considered a fundamental requirement for defining creativity (Boden, 1996; Newell et al., 1963).

> *Creativity is a puzzle, a paradox, some say a mystery. Artists and scientists rarely know how their original ideas came about. They mention intuition, but cannot say how it works. Most psychologists cannot tell us much about it, either. (Boden, 1996)*

Humans are often influenced by previous experiences when creating new things (Veale et.al., 2019). That may be noticed by studying back on the history of art and design. For instance, it is noticeable that the graphic paradigms have been evolving by introducing some novelty on top of influences from the past. Also, it is noticeable a big graphic exchange in the western world due to the introduction of the movable types by Gutenberg around the 1450s, or the introduction of the *Unigrid system* by Massimo Vignelli, in 1977 and later by Meggs & Purvis in 2016. In that sense, it may be reasonable to consider that creativity may be achieved stochastically, for example, by getting to know or getting to use unfamiliar tools and solutions.

**Designing with computational tools**

> *"The field of graphic design has been continuously changing thanks to the impact of innovative technology."* (DePasquale, 2013)

In the second half of the twentieth century, the digital revolution introduced new development tools. Those made possible new ways of exploration, leading to unthinkable graphic solutions (Laurel, 2009; Lupton, 2014). For instance, similarly to what happened with Gutenberg's movable types, also the introduction of computational methods made it possible to create new and unique letterforms — one of the immeasurable possibilities computers provide.

A more recent example sustaining the assumption that new tools promote new creative solutions is the new design era we are living in, made out of dynamic and participative designs (Blauvelt, 2008, Shaughnessy, 2012). The possibility of adding dynamic or interactive dimensions in graphic designs has been opening new paths for novel solutions; For example, concepts may be expressed through animations or brands may be reactive to people or the environment (R2 Design, 2013).

The concept of Artificial Intelligence (AI) was first introduced by Prof. John McCarthy in 1956, referring to the "science and engineering of making intelligent machines, especially intelligent computer programs" (McCarthy, 2007). Until nowadays, the potential of machines to be creative in their own right have been explored by academics and practitioners from diverse disciplines. The concept of Creativity, a widely studied aspect of human behaviour, is now associated with AI, resulting in the concept of Computational Creativity (CC). Creativity is increasingly manifesting itself in different domains, and due to this heterogeneity, the concept becomes difficult to define. Veale et. al. (2019) established CC as "an emerging branch of Artificial Intelligence (AI) that studies and exploits the potential of computers to be more than feature-rich tools, and to act as autonomous creators and co-creators in their own right."

By the combination of creativity with machines/tools appears the concept of a creative system that is, according to several authors (Karimi et al, 2018), an intelligent system that can perform creative tasks alone or in collaboration. We can divide these systems into three main groups, by which the role of computers in creative systems can be characterized: (i) creativity support tools; (ii) co-creative systems; (iii) and fully autonomous systems. The first set, — creativity support tools — are systems/tools for helping humans to experiment and quickly visualize creative solutions, through computational based methods (e.g. Adobe Inc.'s programs). Co-creative systems are those that collaborate with humans in creative tasks. For a system to be considered co-creative it must have at least one interaction between an AI agent and a human; machines and humans must act based on the response of their partners and their conceptualization of creativity. Fully autonomous systems are those designed for generating creative artefacts without human assistance. Human participation ends when the creation of these systems is completed. (Karimi et al, 2018)

These new tools have demonstrated their capabilities for facilitating users to express their creativity (Liapis, 2016). For instance, fully autonomous systems have been successful in fields such as computational art (Colton, 2012; Norton et al., 2013), music (Zacharakis, 2017) or design (Rebelo et al, 2018; Cunha et al, 2018). by applying many different approaches, such as evolutionary processes (Pereira, 2007; Colton, 2012) or machine learning techniques (Heath & Ventura, 2016) for creating fully autonomous, or co-creative systems.

## b. Creative systems outside the design field

Creativity seems to be necessary for the whole work process, not only in the field of design but also in fields such as engineering. Also, the most creativity-demanding phase of the work process is the same between these two work fields — the implementation phase (Robertson & Radcliffe, 2009).

According to the study of Robertson & Radcliffe (2009), engineers are both positively and negatively influenced by the computational tools they use — in the case of their research, creativity support tools. The consequences of using those are argued to be: (i) getting better ability to visualise and communicate ideas with the work team; (ii) technical difficulties to make major changes in the projects as they get more and more complex; and (iii) limited creative possibilities imposed by the constraints of the tools. The authors also referred that computational tools did not seem to be the most propitious environment for idea generation — a propitious environment would be characterised by "large amounts of sketching and discussion" (Robertson & Radcliff). Nevertheless, the use of computational tools may not compete with doing brainstorming by sketching or having discussions; On the contrary, both approaches may be highly complementary, by first sketching or discussing initial ideas and then using computers to explore, visualize and communicate them. For that reason, this last consequence referred by Robertson & Radcliff may not be relevant within this research. On the other hand, the first consequence presented is a benefit coming from the use of a new tool (a computational tool), so it somehow sustains the assumption that new tools are helpful for the creative process. Then, the remaining two consequences make us realise that all tools have their technical limitations which, consequently, limit their creative possibilities.

## c. Enhancing creativity

It has already done some work within the scope of creativity-enhancing research. For instance, Nickerson (1999) presented a framework composed of twelve steps for teaching creativity. Most of those steps were nothing but suggestions for how to study and set up our minds to be more creative. Even though, there were some steps for which co-creative or fully autonomous tools might be used as perfectly fitting solutions; For example: (i) "Provide opportunities for choice and discovery" and (ii) "Teach techniques and strategies for facilitating creative performance".

Also, it has already been done work for enhancing creativity support tools. Shneiderman et. al. (2010) referred to guidelines such as to create tools with "Low threshold, high ceiling, and wide walls", create tools that "Support collaboration", make tools "as simple as possible" and make tools that make it possible to "Iterate, iterate, then iterate again". Also, co-creative and fully autonomous tools may fit in these guidelines, which suggests that the development of these kinds of tools may be a desirable way to go.

## III. METHODOLOGY

Our team of researchers is composed out of three multimedia designers, fluent in designing and programming. That means they often work on creativity using several types of creativity supporting computational tools, from Adobe's software to desktop/web programming, and even physical computing. Having that, the professional experience of the elements of the team could be a case study by itself.

Taking that as an advantage, the team started by meeting and discussing the subject of the experiment. As a result, basing on the state-of-art literature and the self-experience of the team, another relevant question emerged — "the insertion of new tools (computational or not) has always been propitious for enlarging the creative spectrum of people (opening new paths or forcing them to explore in other directions)". If this assumption is true, maybe any type of computational system has some potential to help humans to be more creative — even systems that take no human intervention in the creative process. Thus, it was agreed that the verity of this assumption should be enlightened by studying, not only the utility of computational systems but also the utility of using different not-computational tools.

Also resulting from the team's discussion, it was identified four types of potential creativity-enhancing tools/systems (CETs) to be studied — one englobing any possible non-computational approaches and three referring to computational approaches:

1. Non-computational tools
2. Creative systems/ Computational tools
    a. Creativity support tools
    b. Co-creative systems
    c. Fully autonomous systems

To find the methodology that best fits the goals of our project, an analysis of the literature was made. Grounded Theory (GT) revealed to be an adequate qualitative methodology. For both novices and experienced people, GT is considered a good method for conducting qualitative research efficiently and effectively, because their methods provide a set of strategies for structuring and organizing data-gathering and analysis.

GT has three coding processes: Open, Axial, and Selective Coding. The initial process of GT — Open Coding — involves analysis, comparison and categorization of collected data. In short, it is the process of defining what the data is all about. In Axial Coding, the second process, the big goal is to identify the relationships between categories and subcategories. In the last process — Selective Coding — the identified categories are associated with the core category, ultimately becoming the basis for the Grounded Theory (Babchuk, 1996).

Also from studying the literature, it was realised that a good method for studying the data collected is to make interviews, record them and transcribe them. For this research, it was decided to make face-to-face interviews to understand all the nuances of participants' language and meanings (Charmaz & Belgrave, 2007).

## IV. INTERVIEWS SETUP

The audio-recorded face-to-face interviews were made with 12 designers and computer artists from the University of Coimbra, Portugal (3 women and 9 men, from 26 to 61 years old, with diverse professional and academic backgrounds). According to their background, the respondents could be divided as follows: (i) three design Professors; (ii) two informatics Professors (iii) three PhD students researching on Computational Creativity; (iv) two PhD students researching on Data Visualisation; (v) a PhD student researching on Data Science and (vi) a PhD student researching on Graphic Design. Due to the nature of our research question, it was chosen people who, preferably, have developed work or have been working with people that already used one or more of the previously listed CETs.

The interviews took 15 to 30 minutes and were semi-structured by previously setting up a list of ten open-answer questions that could be slightly changed or skipped to direct the interviews accordingly to the given answers. Not to bias the experiment, the interviews started with generalistic questions.

No more interviews were conducted when the collected data started to be saturated; That is when no new categories were identified and the data became repetitive among interviews. From then on, interviews were sorted and distributed for each element of the research team for the recorded speeches to be transcribed into digital text. After that, the complete set of transcriptions was analysed under the GT three coding processes: open, axial, and selective coding, already explained.

# V. INTERVIEWS ANALYSIS

To respond to our hypothesis, we made interviews, as explained above, divided into three sets according to the theme. The themes' names were presented as follows: (i) Creative process and creativity; (ii) Computational and Non-computational tools and their advantages; and (iii) Fully autonomous and co-creative tools/systems.
By applying an open coding methodology, we were able to identify several categories in the transcripts of the interviews. This way we were able to assess some assumptions to answer our main research question: "How do computational systems influence creativity?". To improve the answer to this question, we formulate three sets of questions already mentioned also understand how (i) computational systems influence the creative process?; (ii) is it worth investing in the development of computational systems to aid creativity?; and (iii) how co-creative systems and/ or fully autonomous systems can be useful in the creative process?
In the first phase of this research, to establish a knowledge basis for this research, we looked for the main influences on the creative process development.

## a. Creative process and creativity

**Questions:**
The interviews started by asking respondents about the stages of their work process and whether those stages imply or not creativity.

*Q1— What is your work-process (design-process) from gathering requirements to final-arts?*
*Q2 — Which phases of your work-process imply creativity?*

By that, it was intended to understand respondents' backgrounds better. For instance, whether or not (i) there were similarities between different backgrounds, and (ii) whether or not the stages of their work-process imply creativity.

**Findings:**
From the first set of questions, it was found that the work process of all the respondents includes the stages of (i) understanding the problem and the project requirements; (ii) searching for what has been done; (iii) combining solutions for coming to a result. Also, it was referred that the prior knowledge and experiences of the person, as well as the way he/she understands the problem is highly influencing the result — the more the person knows, the more creative he/she is. As two respondents argued, creativity is dependent on the context people live in, their knowledge, the books they read and their life experiences.

Although it was consensual that all the stages of the work process may require creativity. For instance, two respondents claim that a simple search implies creativity, not only for the way we search, but also to know in what domains to search. Nevertheless, it was highlighted for two respondents that most of the creative effort regards the implementation stage, and another respondent claimed that the requirements stage is the one that required less creative effort.

It was also assigned that creativeness may come from outside the working process; For example, from nature observation, routine tasks.

## b. Computational and Non-computational tools and their advantages

**Questions:**
The second group of questions were related to the tools computational and non-computational used by the respondents. The questions related to this theme were presented as follows:

*Q3 — Which tools do you use in those phases? Do you usually use computational tools/systems during your creative process? If so, which of these tools/systems best support your creative process? Why?*
*Q4 — From your working-experience, what is the impact of the insertion of new tools in the creative process?*
*Q5 — Regarding the impact on creativity, how do computational tools/systems differ from non-computational tools (pros and cons)?*

We used questions related to tools used by the respondents to understand: (i) what tool of computation or on-computation they usually use more; but also (ii) if these new tools used to help them in their creative process and how; and (iii) which were the differences between computational and non-computational tools and in what context they would use each of them.

**Findings:**
Once we have completed all the interviews related to the use of these new tools we came to some conclusions. Everyone said that they often use computational tools in the creative process. We also notice, during the interviews, that the most commented used types of tools were related to the implementation phase. Besides, part of the interviewed claimed to use creativity support tools on the implementing part during their work process. Tools of organizational support (version control and planning) and search tools are also used. One of the seniors Graphic designers interviewed also claimed to use other computational tools via third parties.

Most respondents said that the great advantages of computational tools were (i) the speed, the (i) time savings in content production and (iii) the great exploration of content, giving rise to alternatives that otherwise would not be thought of — there are new variables in the creative process. Some respondents also claim that thanks to the introduction of computational tools they have no control over the entire process, allowing them to go back and see how the process was progressing. Besides that, nowadays there is a new perspective of content sharing and collaboration in projects due to the easy access to the internet and, as a result, easy access to new tools and support by communities. A respondent also said that computational tools are also useful because they provide a basis for starting. The evolution of the tools leads to the very expansion of the study areas of the professional areas. However, some respondents also declared that thanks to the easy access and use of these tools we became dependents on them to make our creative process; like in the past, the difficult accessibility dictated the limitation of the tools used for a given project.

On the other hand, we realized that, for the most part, respondents continue to use analogue methods (e.g.: books as research; paper and pen as support for quick ideas and nature as inspiration). Throughout the interviews, some respondents claim that, in analogue methods, the whole process of execution and exploitation of the domain is much more weighted and reflected a priori.

During the interviews, it was found that the context, the type and the needs of the project define if a computational system is advantageous or not for the project. A student of PhD working with Computational Creativity also claimed that, sometimes, the combination of computational and non-computational tools can be an asset to generate more experimental and less standard results.

### c. Fully autonomous and co-creative tools/systems

**Questions:**
The final set of questions related to fully autonomous and co-creative tools/systems directly.

*Q6 — Would you use a new co-creative or fully autonomous tool or system to help you with your creative process? What would lead you to use a new tool?*
*Q7 — Have you been inspired by other designers' work who used methods and tools different from yours? If so, do you think they helped you in any way to expand your creativity?*
*Q8 — What do you think about systems of human-machine collaboration, co-creative systems, and the attribution of different degrees of autonomy/decision in the production of artefacts?*
*Q9 — About fully autonomous systems, what do you think about letting a system like this design a book, a poster or a webpage with no human intervention? Do you think these types of systems could be useful in creative processes? Why?*
*Q10 — Do you think that the massification of such systems could threaten human creativity? Or instead, we should use them to evolve our capacities and, consequently, extend existing creative trends?*

By that, it was intended to understand whether or not (i) co-creative or fully autonomous systems can be useful in the creative process (can people inspire in machine's outputs as they do with people's work); (ii) If people would use them in real scenarios; And, finally, whether or not (iii) it is worth to invest on the research and development of such tools.

**Findings:**
It was found that most people think that these systems will never replace human creativity. Instead, we will complement each other because each one is more capable of doing certain kinds of tasks. Moreover, all the respondents expressed their interests in using co-creative or fully autonomous systems during their work process. For example, some of them said that it was due to curiosity, for automatizing ordinary tasks, to access new functionalities and, most of all, said that they use these systems to suggest new solutions. However, there was a clear preference for co-creative systems.

It was also noticed that the interviewed people often take advantage of other people's work to (i) find inspiration (four people); (ii) know what has already been done and criticise it to be able to do things different/better (three people); and (iii) get to know new solutions, techniques or tools (seven people). As one of the senior designers mentioned: *"There are tools that determine a certain kind of solution (...) We need to know certain kinds of tools to get a certain kind of results"*.

These latest answers confirmed that people find other people's work useful for their creative impetus. Having that, to clarify whether fully autonomous systems could be as useful as co-creative systems, we tried to lead people into an analogy between inspiring other people's work and machines' outputs.

From the following questions, it is possible to infer that these systems are indeed capable of helping people to find new creative paths, as well as other people's work. Nevertheless, some have suggested that those systems may be much more effective for helping with objective-evaluation issues.

Also, there were insights about the level of autonomy that these systems should have. In general terms, people defended that the level of autonomy given to the systems must adapt to the needings of each project. Regardless of the level of autonomy, most people hardly defend that the process shall always be guided by human wills. Thus, there were also instances affirming that allowing machines to replace some human creative tasks may not be a totally bad application — people would adapt and direct their capabilities to other more unexplored creative tasks.

After the findings, we applied Morrow & Smith (1995) theoretical model to highlight and to identify relations and dependencies between the multiple categories. The theoretical model mentioned is shown in Figure 1.

# VI. CONCLUSION

One starting conclusion of this research reveals that the creative process is not mainly shaped by the computational tools themselves but rather by the social, ethnographic and personal background knowledge of the user. In other words, we can say that the process is moulded by the way the problem itself is interpreted. Moreover, the use of creativity in the creative process workflow is also consensual, what changes is the how and the where they seek it.

Regarding the impacts of the introduction of new tools in the creative process, well-established evidence is an increase in productivity. This is achieved by providing, in the early stages, novel techniques and/or ways to amplify the exploration and velocity of the production process. Moreover, allowing better control over the state of the whole process thus transmitting higher levels of security and confidence to the user creative process since his process is now more easily revisited, reshaped without affecting the process workflow.

There is also an evident change in creativity and in the mindset of the respondents that seems to be conditioned by the varied use and combination of different types of tools, from manuals to computational tools. These practices lead to new possibilities and therefore new solutions. With these types of approaches, new functionalities emerge due to a pooling of functions and capabilities from different types of sources.

An attempt to question this theme through a more subjective perspective on the role of this type of tools, their application methods and their space/intervention in our future creative processes reveal an awareness of their value, although there is still a reluctance in the full autonomy employed to the machine as a way to reach a final resolution or solution. Instead, there is a consistent view of a creative collaboration between man and machine as the ideal model of creation and use of creativity. In conclusion, to paraphrase one of the respondents, all professions, their process of thinking and execution evolves and mutates in high accordance with the very evolution of their tools. Furthermore, their professionals' knowledge background and past personal experiences have a strong impact on the employment of creativity, namely due to social, cultural and ethnographic reasons.

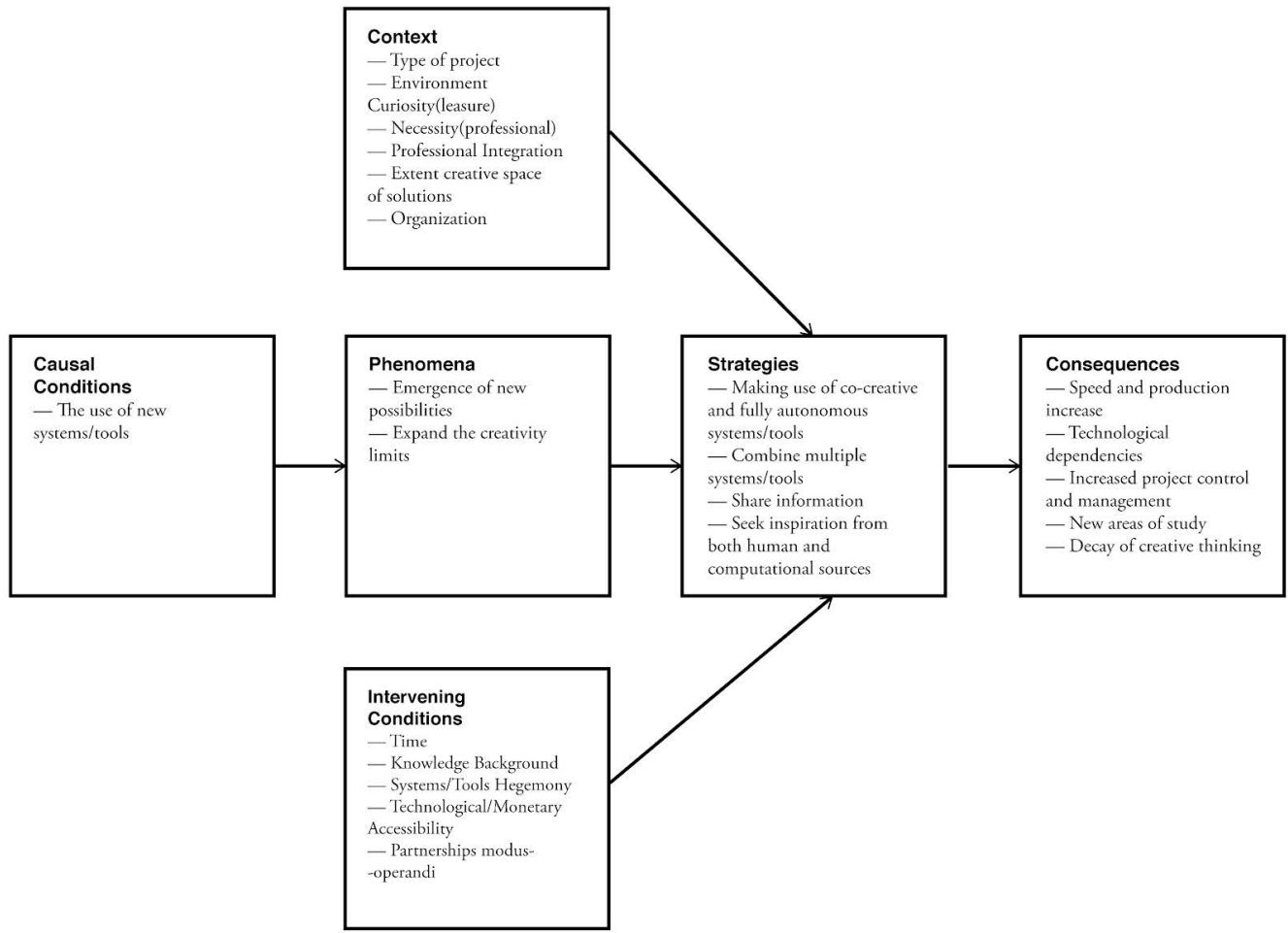

**Figure 1** – Theoretical model of our case

# Attachments
# — Row Interviews and analysis

## Disclaimer:
This is a not-formal/not-official row document that our research team used to organize the analysis of the interviews referred to in the "ICCC 20212 workshop paper, "Performing Creativity with Computational Tools". Also, as these interviews were conducted in Portuguese, something may have been lost in translation.

## Interviewees profiles

All the interviewees study or work in the University of Coimbra (Informatics Dep.) and are directly or indirectly associated with projects relating to Computational Creativity.

**Ph.D. students with design and informatics background (7/12)**
1. Ana (Computational Creativity)
2. Miguel (Computational Creativity)
3. Mariana (Computational Creativity)
4. Catarina (Data Visualization / Aesthetics)
5. Tiago (Data Science)
6. António (Data Visualization)
7. Sérgio (Design Research)

**Professors with informatics background (2/12)**
8. Hugo (Computational Creativity, Poetry, Music)
9. Amílcar (Computational Creativity, Music)

**Professors with design background (3/12)**
10. João
11. Nuno
12. Artur

# A — Row Interviews (Transcripted)

In this section, color codes were used to highlight some concepts that we found relevant to our purposes.

## Q1. What is your work-process (design-process) from gathering requirements to final-arts?

```
Color code:
Identify problem / requirements / questioning of the program
Methodology (related to the first, maybe we can even join)
Draw / create a concept
Search / collection of references
Practical part / create
Documentation
Depends on the project
```

**Ana -** Well, I would say that it depends a lot on the project. But, as a rule, I start by investigating similar projects in the area, to see how they overcame some problems and to inspire me in this to start leading my creative process. Then, I go through several experiments and it is an iterative process.

**Miguel -** First it is to identify what the problem is and then it is to try to design and implement the solution. Going through a process of collecting references, identifying ways to do it, trying to know what already exists and does not exist.

**Mariana -** First I start by thinking about a general concept. If it is a logo, I try to understand what is the message that the client wants to convey, with a poster or a graphic identity. I usually look for images on Pinterest, I make a mood board to give me visual ideas of what the final product might be and then I usually move on to practice. I take Adobe Illustrator and I start doing things, from scratch or based on existing images. There are people who first think a lot on the paper before moving to a more final form on the machine. I do not. I experiment right in the digital medium to see how it would look in the final format until I eventually reach the final product.

**Catarina -** In my current work, information visualization, the data come to me. Then, I define the objectives, sit down and scribble some keywords. Later, through visual forms, I will transform something visual into something that represents data. Then, I research some figures and work in the area to find out what is done, what can be improved and what novelty I can add. Then, it's "brainstorming", mixing it all up.

**Hugo -** When I have an idea I like to quickly check if it works. I like to do a little experimentation to clear up doubts and know if I am going the right way or not, if it is easy or more complex or if it is simply impossible to do. Of course, before I discuss and read what has already been done. I use paper and pen to make some models and I think about the variables to be introduced.

**João -** We assume that the designer starts from an order, although there may also be self-initiated work. In my work process, the more well-defined the order, the better for me. It is not the same for everyone, some designers prefer some vagueness because it apparently gives them more freedom. I like the limitations, I like when the order is very detailed because it is so useful to satisfy those limitations and to work around them. One of the interesting things about the design process is the questioning of the program. To what extent does the designer, as a result of the accumulation of knowledge he has, allow himself to question the program. I think the client knows more about the subject, in general. It is a wrong idea to think that the client does not know anything about the subject. We can know how to formulate the questions better than he does. And therefore, questioning the program is accepted as one of the fundamental processes of the design process. It is something that I use a lot, to interpret and question the program. Then the work is driven by the requirements and therein comes our knowledge, our acquired knowledge, our past practice, our relationship with the themes. Problem-solving is dictated by the problems themselves. There is no ideal solution to solve a design problem. The problem itself and the way the problem is interpreted dictates the solution. Different designers solve the same problem equally well or equally badly with different solutions.

**António -** I try to understand the concepts associated with the requirements and transform these concepts into images. From there, I try to create several types of examples, models. See what works and what doesn't. Trial and error. And according to the client or our taste, in the case of personal projects, I try to understand what works, improve it and arrive at a final solution.

**Tiago -** I start by trying to understand what are the goals that I have to achieve. Then, I try to see what I have at my disposal, what conditions I have in terms of hardware and time. At the same time or later, I see references, works, and do research. I try to find explorations that, in a certain way, try to solve one or part of the goals that I identified. After identifying some projects that I find interesting, I try to understand what technologies are used, what these approaches allow me to do and what are the disadvantages of these. I try to understand if it makes sense to do something similar or if it is just a starting point and see how it fits with the state of the art. Then, there is a phase of development, where, typically, I create a version and develop little things, trying, at the same time, to see if they work in conjunction. When I see that all the goals that I set at the beginning are accomplished it means that I have reached a final version. In the end, I make the corresponding documentation.

**Sérgio -** I usually work a lot on trial-and-error. So I like evolutionary algorithms because they work a lot in the same way that I work. In other words, you start from scratch, you have an idea, you have something you start trying and keep working until it is really something you like. Sometimes you also have these problems of local maximums, when you reach a point where you stop, you are blocked. As I am very unstable, one day I can take one approach, the next take a completely different one. My process is very "broken", iterative and cumulative until I reach something that satisfies me.

**Nuno -** Whenever there is a request or I identify something that will give rise to a project, usually, what I do is a phenomenological approach. Previously, I started by having a clear idea of what I am going to do and then I follow that mental idea a little. I leave the formalization completely open because the fieldwork itself is going to say what I am going to do next.

Usually, I try to approach the object, think about the object that I will have to formalize and, in this case, get to know the entity that is asking me for the job in its different variations. Then, I think about a project that suits the customer. I usually say that a creative process is half authored by the designer, half authored by the client because I'm not going to draw anything in the void. There is always a question that is asked from the outside and then my design project is in response to that question. I have always worked very closely with the client and made decisions with him. He was always aware of all the stages of the process and he always had to approve the process. Then, later,

I started to work regularly with the same people, because there was already a little bit of empathy in terms of work. There, it was easier to have this validation.

**[our team] So about the process ...**

I take a phenomenological approach. I will see who the customer is, with whom the customer transacts. I will get to know the space, meet the people, etc. Then, I try to understand the project dimensions, and then I adjust the project for that. First I gather knowledge and then I let the real information that exists on the ground tell me the solution of the project (the conversations I have with the client, the site visit, etc.). Let's say it is an approximation to anthropological methods.

**(this part does not regard the issue of this question.
Can be applied as an answer to further questions)**
In a creative process I see the tools in a very secondary way because, normally, I also chose which were the most suitable tools according to the project. Usually, as a teacher, I always see this issue being raised in class. Students feel very much the need to learn how to handle the tools. Usually, from the students' perspective, students always think that the knowledge they are going to acquire in universities is a purely technical thing, and they, knowing the technique, can develop any type of project. And it is a lie, it is to say that anyone who has a computer can be a designer. It is not, it is not having access to the tools that, suddenly, I can manage. If you put me in front of an airplane, I can't take off.

Design is more a discipline of analysis, criticism, observation and then decision making. I prefer to teach people to think. The tools come after this whole process, that is, I will choose the tools I want to work on, because, many times, I do a design project that doesn't even involve computers, because, for me, a computer is more of a tool among others.

Obviously, nowadays, it is impossible to avoid this, because there is always a digital process in my work processes. But I don't want this technology to be a condition for the formalization of my ideas. I want tools to be at the service of ideas and never the other way around.

When I started teaching, here [University of Coimbra] ten years ago (I came from Fine Arts), (this is going to be controversial) I doubted this logic a little, being here creating a course in Computer Engineering, half in Design. I felt, very sincerely, that this course was a little ahead of me, it was already a step ahead of me. And it took me a long time to fully enter into this logic, and now I have been a high advocate for this logic for many years. Exactly, because that's where things go. Because, today, we consume more information from digital devices and things are more and more going in this direction. So it is natural that, in other words, just like in the course, I had to learn what printing techniques were so that I could adapt my design project (ex: knowing that this is technically impossible to do, or very difficult, or this will make work twice as expensive) obviously, you have to understand the code very well, to understand how you are going to design solutions for dynamic supports, which are not static — they are always changing.

Technical evolution has to do with the history of design. I mean, also in the '80s when the computer was introduced in the processes of making design, there may have been a rupture, but I see a continuity in that rupture. It accelerated the process of streamlining ideas, because suddenly, instead of making a model in two days, we can make five models in one hour. It made the execution of ideas a little faster and greatly shortened the time to make a project.

**Amílcar -** When I think about the creative process, I have to think about musical composition, which is where I put it into practice. The collection of requirements extends over time. My requirements gathering involves reading, discussing... I watch the rehearsals to understand the context, collect it myself and try to find out what the director (client) is waiting for. The requirements are imposed by him and to these requirements I impose my own and start to set aside hypotheses. My goal is to try to converge on something. In parallel with that, what I usually do what I already do on a daily basis, but now in a more directed way… it is to listen to a lot of music (see the state of the art). I collect inspirations about environments, certain forms of melody and then I start using the computer. For me, the computer is the basic model. It is with the computer, sometimes with the help of a guitar or with my own voice, that I begin to "draw" the sketches. I listen and "draw" without showing it to anyone, for some time. So far I think it is ready to show to the director (client). Usually in this process, I go back and redo new versions, etc. The model is being structured and getting more and more complete until the point is that I put that in place with the actors and it is the first proof of concept. This whole layout process and all the adjustments from there are done on the computer. If I didn't have that tool, forget it... only if I dedicated myself full time.

I don't have a requirements analysis phase, they are always being collected. I saw how the audience reacted and I will adapt.

**Artur -** The design processes vary. But I always worked together with another designer, and, since college, we have always been used to talking a lot. I think that the beginning of a design project involves a conversation, the reading of the program, if there is one. If it is a book, it involves reading the text, discussing it with the various stakeholders in the project. After we have this information on our side, we move on to a list of key ideas, the key ideas can be transformed into a scheme and a scheme can be transformed into drawings.

As we work a lot with typography - typography is our work tool -, the first materialization of ideas goes through typographic composition. Usually, the execution of a book, a poster, a website, or an identity always involves written text. So we usually test different fonts, different scales, draw fonts from scratch, combine fonts. Then, we tested several clues, eliminate some and, sometimes, we recycled some that we had already eliminated and confronted. When we come up with a finer idea we start to refine it. There is always a look at the things that we set up in the past, during the process, to see if there is anything that can be used, and then, when it is shown to the customer, this interaction process is often resumed. The customer's opinion creates a possible deviation that can bring back the first clues.

I am summarizing the process a bit, but most of the time it is like this.

# Q2. Which phases of your work-process imply creativity?

```
Color code:
I find creativity everywhere
Phases where there is creativity
Interesting topics
```

**Ana -** All, practically all. From (i) the research phase, that is, the type of research that is done (getting out of the box a bit, sometimes) to obtain certain results in the search for similar works; (ii) the construction of the tool itself; (iii) in the way the problem is approached, mainly there... How will I find a solution to my problem?

**Miguel -** I think creativity is everywhere. There is not a stage where I say "I'm going to use creativity here". Creativity ends up happening when I least expect it, when I'm thinking about anything is when I'm more creative (running, everyday acts). There is part of the process where I strive especially to try having ideas, but it is not always in these phases that ideas appear.

**Mariana** - Deep down, in all of them, but more when you already have at least one concept and you have researched a set of images and steps to apply that information and inspiration to the final image of the work. (when you're already creating)

**Catarina -** All, even to do research it takes creativity. Sometimes we have to do more specific research and we need to think a little to get to the keywords, and even that requires some creativity.

**Hugo -** There is always a part of creativity right at the beginning when you have ideas, brainstorming or devising the best solution for a given problem. You have to be creative, of course, at the limit, you can do what others have already done with a small change. Especially when in collaboration with other people you can take advantage of what your work has in common or see what can be used from each of them to obtain a better solution. I believe that this can be considered creativity, in a way.

**João -** In this way of looking at things, it has a lot to do with the stage of understanding and questioning the program. He often dictates the solution. When we think more deeply about it. To understand the problem, we have to investigate the matter. In addition to the issue of creativity, it is something that I consider very atonic, it is the subject that we have to deal with a few times is about design. The subjects are different. We, in the end, are always dealing with matters that are not design. And we solve problems better the better we know them. If by chance designers with more culture and openness to other aspects than just design aspects (everything that happens around them, in society, in politics, in culture in general), in general, they solve the problems better. Sometimes I think it is not so much a matter of creativity in itself, I believe the better you know the subject the more creative you are. This can happen in this phase of questioning and investigation of the problem but also when selecting the medium to work on. We may have the problem very well questioned and the solution found and then find it difficult to find a materialization for that. The medium we choose can often be dictated by the novelty of new means of dissemination, and may often not be the most appropriate, which also leads to being creative in choosing the best dissemination approach (Example of the CD when I was in Vouga. All of his customers wanted to use it but were often not exactly the best solution to the problem).

**António -** The whole process except the evaluation. The beginning, the understanding of the concepts, how to explore them visually (probably the part that implies more creativity) and also the combination of these concepts figuratively. Therefore, in the identification of concepts and their combination to generate new ideas.

**Tiago -** The phase that has **less creativity is the requirements part**, there is no need for any creativity, we have to be very objective. On the other hand, even simple research requires creativity. You have to be creative not only in the way you look but also in where and in what domains. Of course, it also takes creativity in how you are going to tackle the problem, the approach. Creativity is more present in research and in the way in which the problem is tackled.

**Sérgio -** All of them end up implying the same level of creativity that is being accumulated throughout the process. In my opinion, **creativity is not a gift but the result of work.** The technique, the general knowledge and the way you work overlap with creativity, or rather, expand creativity. Through past practices, knowledge accumulates that influences future practices.

**Nuno -** I think that creativity is in everything, even in research methods. In other words, innovation in methods, before carrying out the project, can also help improve results. There is this very rigorous scientific method where you first draw a methodological plan and then go to the field to apply and check tasks, however, I think that it is not a method that fits well with the design. I have never replicated the same process, each project is a project. Knowing how to redesign all methods and objectives, client by client, project by project is something that only a creative person can do. For this reason, I do not see creativity in the formalization of the final object, I see creativity in the whole process (eg, in decision-making, in research).

**Amílcar -** All. From the beginning, requirements analysis is in itself a creative task, unless we just accept the requirements as an order that we have to respect. But for me it never happens, because to the requirements they give me I add a few more. Choosing these requirements themselves is a creative process that will impact everything I do. From requirements to the design of the solution, everything is a creative task. All phases have a creative side, if we want.

**Artur -** I think that creativity is always present in all processes. It is even difficult to define what creativity is. I think that when there is not a strong idea, a strong concept, what we do is to draw - to provoke our own creativity. For that, we can put into practice exercises that have to do with the content and the form, print them or compare them. Then my work partner and I analyze, comment and discuss. Then, I think there are good ideas that come up looking at printed materials - the way the letters touch each other, how a shape can touch a letter or how a shape can turn into a letter or vice versa. It is such a complex web that we can create atlases of images around our own production and the references we have and those given by the project itself. For example, if it is an architectural project, the forms of the elevation and the city where the building is inserted can be transformed into ideas. They can be more or less abstract ideas and this web is being assembled, sometimes the difficult thing is to narrow it down. **In graphic design, sometimes, the solution is not a lot of ideas that confront each other, but it's something that sums it up.**

[The interviewees always answered the questions 3 4 and 5 by answering to other questions]

# Q3. Which tools do you use in those phases? Do you usually use computational tools / systems during your creative process? If so, which of these tools / systems best support your creative process? Why?

```
Color code:
Used tools
Partnerships
Why using tools
```

**Ana -** I use <span style="color:red">computational tools</span> or <span style="color:red">tools for collecting information, for organizing (eg: Dropmark, collection of visual information) Processing, MaxMSP, Arduino,</span> always depends on the project.

**Miguel -** It depends on the type of the project, it depends on the type of tools I have in hand. I usually use <span style="color:red">research</span> to try to identify, using keywords, some kind of visual ideas or concepts that interest me, but it depends entirely on the project.
…
My doctoral thesis focuses on that. I am currently trying to develop a system that helps me to foster creativity and try to create a system that the user can interact with to get more ideas, which he may not have at the moment. And so, that ends up being my goal, trying to develop my own tools that allow me to foster my creativity.

**Mariana -** <span style="color:red">Search: Google. Pinterest.</span>
<span style="color:red">In freelance graphic design jobs: Adobe programs (Illustrator, Indesign)</span>
<span style="color:red">Computational: Processing and MaxMSP.</span> I use them to help me organize the data I need and then pass it on to another type of data (musical data). Data organization.

**Catarina** - Internet. Some books if it is something very specific. Sometimes being closed is not good, the street or other places can also inspire you

*[our team] And in the development process, what tools do you use to create?*
Pencil and pen, a piece of paper and then I move on to implementation.

*[our team] and in the implementation, what tools do you use to create?*
Computational: Java, or javascript.

**Hugo -** In the initial process, read and write. Writing helps a lot. It is also important to talk to other people who have read the same things or who are in the same field so that they can help you come up with new ideas.
Computational: Information management and sharing tools. For example the <span style="color:red">electronic organizer</span>.

**João -** I mainly did editorial design and always used the tools, by "obligation", more accepted in the industry and in the market, such as <span style="color:red">Adobe Indesign, Illustrator and Photoshop</span>. <span style="color:green">I rarely use alternative tools to keep from being integrated in the market and to stop having correspondence with other professionals, products. There is a certain hegemony in Adobe Studio that dictates this issue a lot.</span>
As an art director, I work with much more varied tools. <span style="color:orange">I work with the tools that those who work with me work with.</span> In motion, <span style="color:red">After Effects, space equipment, CAD, 3D renders, or even programming tools</span>.

**António -** At the beginning, writing (paper and pen). Description of the ideas, what the problem is, outline objectives. Depending on the type of project, in programming projects: Processing, eclipse (code editors). In image projects: Photoshop, Illustrator, Indesign.

**Tiago -** Development software: code editors (example: IntelliJ); version control software (eg. git) and writing tools (eg. latex).

**Sérgio -** For graphic design projects, which involve drawing and composition, I usually use Indesign and Illustrator. It depends a lot on the context. For graphic identities, I also resort to Processing to work the dynamic part or to generate several results in a very demotivated way of beauty. For example, I used a minimalist interface just to help evaluate results, using web programming (javascript). In other more technical projects or evolutionary projects, I use python.

**Nuno -** Everything. Paper, pencil, computer, search engines, books. That is, everything. The project itself dictates where I go to get information. Invariably, as I am now doing a lot of projects around researching the past - history of design or history of design related to other things - I use its archives, the internet... Search engines are interesting, but to see what happens. For example, I remember that for a logo, a person will always search to understand what is related to that word, to make a survey of the state of the art, to understand what not to do, not to commit involuntary plagiarism.

*[out team] And in relation to computer systems? More to create.*

It depends on the project. There are tools that I don't know how to use and for that reason I always have to resort to people from outside. Then, I become a kind of art director and explain to the person what my idea is. That's why I say that it is also not very important that I have to know how to work with all the tools. I don't know how to write a line of code, but I know how to communicate with a programmer so that he executes a certain idea for me. The important thing is that I know how to communicate with him.

**Amílcar -** The tool [talking about the tool he uses to compose] is the key to everything I do. In addition to that, I also use other tools (algorithmic tools) that produce a rhythmic basis in a certain style. It also helps me to look for already existing patterns of a certain style, to join and then adapt. All these adjustments, all these refining cycles, I can only do them because I have a very good tool, with which I can work very well because I have been using it for many years. [response to Q1]

**Arthur -** It's variable. We draw a lot on paper. We use photography, we draw on top of photographs, we take notes, but we also use a lot of computers. The computer serves us, sometimes, to do the editorial assembly - to bring together all these drawings and contents and photographs and our own productions.

Sometimes we draw by hand, but most of the time we draw on a computer. Then, those images that are generated go back to the computer to make an assembly, as if it were a narrative. We create a sequence of images and often print them to confront them differently. When I talk about assembly, I'm talking, for example, about creating a sequence of images in InDesign and organizing those images in a logical way - a temporal sequence that allows us to go back and understand how evolution was. In addition, we are also interested in printing it all and shuffling the sequence that is proposed by InDesign.

# Q4. From your working-experience, what is the impact of the insertion of new tools in the creative process? (If you have already partnered and felt the benefits of introducing these tools, even if not directly, you can talk about it)

```
Color code:
Impact of new tools
Interesting topics
```

**Ana -** It allows me to expand the results, to go further in my experimentation.

**Miguel -** I do not have concrete information at this moment that will allow me to say with certainty whether they do help or not, but my idea is that these tools do help a lot and do have the potential to help a person who is in a "creative block" overcoming this situation. The computer system allows you to come up with ideas that you were not having at the time... or through ideas that you have at the time, reaching other new ideas that you would not remember. So I think that there is great potential to be explored.

**Mariana -** It's spectacular! First, you have many more possibilities, mainly in computational, generative systems, generating animated artifacts which transform themselves. And I think it helps a lot more in the creative process, at least for me, because I feel that I can sooner try joins, combinations and check if it works or not... what paths I should or should not follow, allowing the process to be much faster. But I've also grown up with them and I don't know the differences of designing without them.

**Catarina -** I think it helps a lot. Especially if you think about graphic design, in patterns and in the composition of posters, now you have the facility to go back and be all right. Before, if you were drawing by hand and if you were wrong you had to start again. [answer to the next question]

**Hugo -** Yes, they increase productivity. On the other hand, they also make us more likely to be attached to it, since when they fail, the process also fails. At the same time, they make everything that is done more intense. There is not much time to rest or not think about things a bit because everything is within reach.

**João -** In my story, it was "brutal". My story is in the books. What we talked about in contemporary history, starting in the 1980s, I experienced this transformation. The appearance of Desktop Publishing... I did work with other old processes, such as photomechanical processes, photo composition, enlargements / reductions, lightbox assemblies, but not so much with linotype and lead characters. For me, the appearance of the Macintosh was radical. I experienced all this evolution, from the first tools, black and white screen, first laser printers, Page Maker, Apple's connection with Aldus and Adobe, the appearance of the first pagination tools... It is absolutely radical. (He gave an example of working with layers, seen in the documentary "Helvetica"). Making a poster with 2, 3 layers was "hell". With the advent of the Mac, the possibility of changing the font with 1 click, changing the font size with 1 click, changing the color with 1 click was something surreal. It hasn't changed the design in terms of quality, but as Wim Crouwel says, it has added speed. We are much freer to create because the technical processes were terrible, so "heavy", that it did not leave much time and energy for the creation itself. Computing has brought tremendous speed to the process.

**António -** In terms of programming tools, since I use it most, it is mainly Processing. It is very interesting to apply programming to imagery because we are creating a whole system behind that influences that same imagery. We can create an algorithm to structure and define the properties and mechanics associated with the visual elements and leave, independently, the imagery, design and aesthetics of the project to the user. The tools allow the separation of these two issues without being dependent on each other, much less a problem for the user.

**Tiago -** (We are already a bit on the crest of the wave, but we always catch the wave) I don't know how it was before, but I notice the development, a gradient, which is marked not only by the appearance of the software but also by what comes with the software, in the community. We have the classic Processing. Processing when it appeared in my work process didn't just add another program to the computer. It was a software that allowed me to generate visual and interactive outputs in a quick way, but it also brought a community. I think that is also a great innovation. It brought together thousands of people and helped them in their process, including me. When we try to solve a problem, in most cases, we no longer start from 0, we start from someone's point of arrival, expanding the work of other people. With the advent of Processing there is much more information available through, for example, communities.

**Sérgio -** The evolution of graphic design is intrinsically linked to tools. Ever since, in the industrial revolution with the introduction of new tools, mass production, forms of printing... The same in the 1980s with the introduction of the personal computer and desktop publishing. The same thing now, with the current tools, dynamic identities, responsive web design, etc. appears and practices change too. New design problems are created, new spaces to explore leading to the growth of the profession. The whole profession grows dependent on the tools. In our case, it will explode with artificial intelligence, for example. What made the limits of typography expand (in the 90s) was the experimentation made possible by the computer. It was possible to work on elements such as non-linear numbers, tabular numbers, distances and other special characters that were not previously due to the technologies of the time (e.g. mobile types or photo composition). Technology has enabled the use of "lost" elements. There was no typographic evolution, but there was an evolution in the possibilities of its application.

**Nuno -** It is the speed with which ideas are executed. But this has advantages and disadvantages, obviously. It has advantages because we were able to make models in a single day when it was previously unthinkable. (In fact, I had a Professor in college who did not adhere to computers, and he said that doing work on computers was cheating - it was a scalpel square ruler, decal letters). But the funny thing is that when we were forced to make a model with these analog tools, we forced ourselves to think hard before we spent the next three hours making a model. There was a reflection on what we wanted to do. Nowadays, changing the color is automatic and, suddenly, we make these models and then choose the one that looks best. If I had to choose between one and the other, obviously, I would choose the digital process because it saves time. But I think that the ability to reflect has been lost a little, because it is the computer that shows us whether it looks better in green, red or blue and, suddenly, I choose because the computer shows us, instead of trying to imagine it. Trying to imagine forces us to think more.

Due to the fact that the time is much faster, there is a greater demand for deadlines. The deadlines are even shorter on the part of the client, because the client thinks these are automated things…

**[our team] Yeah, you click enter and it's done…**

But this has to do not with design, nor with computers in design. It has to do with the introduction of mobile phones and computers in everyday life. The notion of time has changed a lot and I think for the worse because, culturally, we are losing things like patience, thoughtfulness and we are valuing a lot the immediacy.

**Amílcar -** There is some uniformity, this is evident. It is easier to start, in music for example, to detect that there is a set of songs that follow more or less the same scheme, more or less the same basis. It has become very easy to do things online, things in mass. Anyone who wants to produce at high speed has lots of solutions for making music in the style of x, by copying and pasting. The use of computational tools to accelerate and to facilitate access promotes a lack of diversity, in general. On the other hand, when we give access to more people who would not otherwise have come together, we also have bringing different ideas and exploring things. There are things that I explore and that if it weren't for the computational tool I would never think to do. There must be a lot of people who set out to try more different things because they have a tool that makes life easier. If there is more access, the risk of having worse musicians is greater, but we can also have better musicians. Basically, we are reproducing what humans would already do, but on a different scale.

**Artur -** As a matter of fact, we felt the impact because we were the generation of the transition. Although the Mac already existed when we were studying, we did not have easy access to the computer (I shared a PC with my work partner). In college there were four computers for all students and the students from the fourth and fifth year had priority. The access to technologies was not easy. Because of that, there was a phase when we used a lot of the letters to print, the photocopy, the photograph, the illustration and we got used to assembling things by hand. The computer, for us, was extremely important because it allowed us to work in a much more rigorous way, much faster.

We also realized that, due to the fact that we experienced that moment of the technological revolution, many designers of our generation believed that the computer had appeared, simply, to replace manuality, in a more perfect and faster way. For example, filling in an area of black (which used to be made with gouache or china ink) has become a quick vectorial thing. I remember João Machado's posters that had flat surfaces... he had an art-finalist who cut Pantone colors with a ruler and a schizato and assembled those sequences of flat, beautiful color images that made those posters very impactful, because the color surfaces were very large, but everything was done by hand. Gradients, in the early nineties, were made using airbrushes (I can't swear, but I think João Machado also used airbrushes) and so the difference was not very noticeable when he switched to the computer, because it just speeded up the work process. If you see a poster of it before and after the computer, you don't understand which one is done on the computer.

But there were a number of designers who took the aesthetics of the computer in their favor. I am remembering April Greiman, who took the pixel as an aesthetic factor that was part of her language. I saw the old fashioned thinking that the computer came to help the process and the work continued to be the same, while there were a number of designers who believed that the computer could produce its own aesthetic and we could take advantage of it (as happened with the letterpress, where printers and designers took advantage of the aesthetics of printed wood).

# Q5. Regarding the impact on creativity, how do you think that computational tools / systems differ from non-computational tools / systems, that is, manuals (pros and cons)?

```
Color code:
pos and cons - analog
pos and cons - digital
what is more valuable? It depends
the two are potentiated
```

**Ana -** Analog experimentation always allows us to manipulate the real world more, to manipulate data that comes from the real world to the digital world in a more natural way. Digital experimentation allows us to do a series of experiments more quickly and, perhaps, arrive at solutions that we had not thought about or suggestions for solutions that may be useful for our problem and that had not occurred to us. On one hand, we have a more palpable manipulation, on the other, a manipulation more based on expansion, on experimentation.
In addition, it is also possible to integrate the two – the results can be quite good. There are then several options: (i) only analog experimentation; (ii) digital-only and (iii) the integration of the two. The three approaches will always give completely different results and the application of each one always depends on the project, the context.

**Miguel -** It totally depends on the project. Regarding analog tools, I just remembered those letters with phrases telling us to do a certain type of task. Even musicians use this to try to create different compositions using different instruments or playing them in other ways. Taking these as analog and a computer system as digital (not analog), I think that both have advantages.

*[our team] In relation to the production of an artifact, how creativity is affected or guided by making the artifact in a computational or in a manual (non-computational) way?*

I think it depends on what the person is most used to. Maybe a new digital experience is more advantageous to a person more accustomed to analog than for me, as I am more accustomed to digital. And if I go to the analog, I may see things there that perhaps a person from the countryside does not see **[Miguel is from the city]**. I think it depends a little on a person's normal work method. What you experience, what a person does and ends up having ideas which he likes depends a lot on what they are used to.

**Mariana -** It depends because the good design was already done without computers. As they only had paper, they were able to explore much more of a series of formats than today. Perhaps, we no longer explore as much. Perhaps now we explore more the transformation of the shapes digitally. I think it's a different spectrum, without necessarily being better or worse.

*[our team] What were the advantages of choosing other types of tools, in this case from digital to analog / manual?*

It depends on the type of work, concept... If we want to look for other qualities that in other tools are more difficult to achieve, for example, the organic quality of hand tools.

**Catarina -** In terms of research, we can now use the internet, whereas in the past we had to use books, magazines... Access to books could not be for everyone, it was not so easy, whereas now with the internet anyone has access to these. Maybe there are more people having access to the internet than there were people having books and more specific books in each area. In terms of ease of access, the process has been greatly improved.

*[our team] And about the creation?*

It depends on how connected we are to the need of seeing references. It depends on the person. If someone is creative on their own they may not feel the impact so much.

*[our team] more about the tools: analogical vs digital?*

With these new tools, it is easy to go back, which was not possible before.

**Hugo -** Focusing on the part of the process that involves writing, previously I used paper to store information, organized in dossiers. Now, with digital tools it is easier to have everything collected and organized in a space, it is more difficult to get lost. They facilitate and streamline the content production and organization process.

**João -** This speed introduced by the computer has become, in some way, also harmful because starting by immediately drawing on the screen suggests an increased resolution of the problem, with a more rigorous and final aspect, because we stop using the sketch. When we do this process manually, when we work on paper, with a drawing or sketch, the result has no final aspect. We know that it will go through another phase, another process with more rigor. When we draw directly on the computer, a lot of the reflection and a certain slowness that was favorable to that reflection is lost. The speed can be bad, although in the meantime this aspect of the process has been diluted because our relationship with the machine is also no longer a fascinum due to its speed. The machine today is no longer just speed because we no longer have the other reference passed as a comparison term. [He stopped writing by hand and resorting to sketching. He does it now using the machine].
...
Making a poster with 2, 3 layers was "hell". The technical processes were terrible, so "heavy" that it did not allow much time and energy for the creation. It hasn't changed the design in terms of quality, but as Wim Crouwel says, it has added speed. Computing has brought tremendous speed to the process. [answer from the previous Q]

**António -** Computational tools are already prepared to deal with the main types of problems that the users seek to solve. Older tools have more benefits for more casual users, less experienced due to their easy learning.

**Tiago -** In the initial phase, it is easier to diverge on paper, you want to have diversity. Making a layout on paper adds a layer of difficulty. At the same time, at a stage when you already know what you want to do, when you go towards the specific idea, there are many software that helps to make the idea more concrete, more real (ex: Illustrator).

**Sérgio -** There is the romantic idea that in the past it was good. I think it is not worth fighting against evolution. Everything, sooner or later, will be automated in design. With the introduction of the computer, there are no more art-finalists. I think there will be positions that will disappear, in the same way, in the future. At the same time, they can **open space to explore other fields in other areas of study.**

The analog will remain as an exploratory method, or only as a last resort, in most cases. Although the use of analog processes as a provisional stage before the digital stage will always happen. Digital will be the basis of all work, analog will be the niche for people who like to master processes, but never isolated from digital processes.

**Nuno -** [answered in another Q]

**Amílcar -** There are circumstances in which non-computational approaches have a lot of advantages. However, when a group of people gets together in a cafe to play, what is there in the first place is the social relationship and the people playing together. Perhaps a computer system is not a substitute, it is not essential for people to have fun and play. Technology can be important in situations where society is not so critical, the composer who is alone. The musician wants to be prepared for when he plays with others and can use a computer system that will allow him to practice for a longer time without having to have the others there. I think that one thing does not take away the other, and I think the computer has a different impact when you are alone and it loses a lot of use when you are playing together. [not exactly what we intended in this Q]

*[our team] In relation to the process of creating?*

It doesn't mean that we couldn't create without the computer. I could sit at home with my guitar and invent things, but I could not take them to a point where I can take them right now (take them to a theater play, alone, without needing to be bothering others) that would not be possible.

There are also circumstances in which the computer can only get in the way, namely when the goal is to play together. It is always complicated to introduce tools with some automation, for example, because the group's pulse is felt with people looking at each other and listening to each other. There will be a time when the machine will succeed, but until then, you still have some difficulty playing together. [not exactly what we intend in this Q]

**Artur -** I can't identify just now a pro and a con. I only see the advantages of having both. We have the ability to think in an analog and digital way, one empowers the other. There are certain things that have been produced in an analogous way that have been influenced by the digital and vice versa. Of course, digital has transformed the lives of designers and everyone else. Now, designers produce much more quickly, for much more diverse media using the tools of computing, without a doubt.

## Q6. Would you use a new co-creative or fully autonomous tool/system to help you with your creative process? What would lead you to use a new tool?

```
Color code:
Would use a tool like that
Reason to use it
```

**Ana -** If tools that I don't have do come to me, it's always good. More important than the tool is your methodology; the tool is a means to an end and what brings great advantage here is knowing how to think, knowing how to have a method.

**Miguel -** You have 2 possible answers, both valid. One is regarding the necessity and the other curiosity. You may need to switch to a new tool either because previous tools have stopped working or because you see advantages in it and therefore assume it adds value.

*[our team] What advantages do you associate?*

Imagine that we are talking about making a poster. I identify that program X has more advantages or offers more features than program Y. I think that anyone who reaches this conclusion goes with the tool that seems to him it is the best. Then there is another side, curiosity. If you end up identifying a certain type of technology that arouses your curiosity and you want to experiment without knowing if it will bring you advantages or not, then I see advantages. If it brings advantages, curiosity can turn into need, influencing each other.

**Mariana -** Yes, the unexpected. You don't know what's going to come out of that. I think that is its advantage. You don't know how the tool will generate things, by itself or using a starting input, or how it will grow considering that its goal is not known. Although, I always find it interesting, at a certain point, to pass a new input in the meantime so I can guide the tool, but always allowing it to grow and evolve within the intended path.

**Catarina -** It is interesting just out of curiosity, things that we never thought can come out and may even make sense, but I think I would not stop having my process. I could do both in parallel and in the end I could compare it as if it were a partnership. However, I would never replace human work, unless I had a lot of confidence about what would come out of there and that it was just great, but that way I would be needed either. I'd use it more for suggesting.

**Hugo -** [He gave example of the "Co-PoeTryMe: Interactive poetry generation" (see article: https://reader.elsevier.com/reader/sd/pii/S138904171730311X?token=A13425698DA99CCF133BBC91414E61346934D9FB575494BC0FC8929474C6F79D0D9FE96900BF5AB8E55006A1F6F5B2D3)].

Most of the time people are not satisfied with the final product produced by the system. They wanted to change the final result, which led to the creation of a co-creative version of the same project, which through an interface allowed the user to manipulate results in a variety of ways. The system could produce a root poem and the human could later change it, or the human writes the poem and asks the system for suggestions or even a combined creation. If I would use this system? I think so. Although not 100% automatically in this specific case. Musicians and poets were an inspiration for the creation of this type of co-creative tool because they said it was an extra help in the production of their content.

**João -** Yes, although I personally do not use these. I do work with people who make use of these tools for their obvious advantages. I see a parallel between my experience of breaking analog methods and moving into digital/computational support, with the latest ways of drawing using programming. This type of tool has two qualities: (i) support for tools that already existed, providing new solutions that the tool alone cannot solve, such as certain automatisms (Ex: Indesign scripts) (ii) and making it possible to draw entirely with programming, resorting to randomness, surprise, error... All of these reasons make the use of these tools very interesting.

**António -** It could be interesting in the creative process, although not totally autonomous in the whole process, but perhaps in its initial phase to create ideas. But yes, I would use it at an early stage.

**Tiago -** My motivation in developing and using systems like these has to do with the possibility that, in an initial phase, very quickly and without much effort, one can create a vast and diverse set of solutions to the problem (many of them may not even make sense). The automatic generation software does open the horizon, showing solutions that you would never imagine. After the system shows these possibilities, I stop using the autonomous software because I already had some ideas shown without much effort. Thanks to this, I save energy that I can spend on the most detailed versions of one of several solutions.

**Sérgio -** Exploration of the field of work, at the beginning of the projects. With these tools we can prototype thousands of examples and then, based on the results obtained, choose the best options, taking into account many other possible approaches. This advantage of presenting results circumvents situations in which we choose an approach, bet on it until the end and then see that it does not work or is not feasible (local maximums). These types of tools are very important to explore the creative process in its early stages.

**Nuno -** The question of automating certain technological solutions is already influencing my thinking. For example, when a client comes to me or when I do a project, I am already imagining dynamic things - it is already natural in my work process.

*[out team] Why did you decide to use such tools?*

Nowadays dynamic supports are no longer ubiquitous, they are everywhere. For example, we are at a time when identities will be seen more often on a dynamic medium than on a static medium, therefore, the dynamic question of that identity, not being a monoblock, is highly valued. Design nowadays follows this evolution. So, as I am a contemporary designer, I also follow the advances. I insert myself in this medium, therefore, I also start to think that way.

In addition, the formalization of a design project does not live without support. But never confuse support with the design project. There is an evolution on the supports, because technology advances, therefore, the design adapts itself to those supports in which it will be formalized. We are now talking about the dynamic issue, but at the beginning of the twentieth century we were also talking about how color suddenly appeared in the printing system. It was a revolution and designers also had to adapt. I think the issue of screens is more of a support. There are going to be things that, at the moment, we cannot imagine.

**Amílcar -** I never use totally autonomous because I want to have a role, the fun is to participate in the creative process. There are circumstances in which I really like having the machine with some autonomy, either to prepare ideas that I have in mind, or eventually to force myself to have alternative ideas. The person himself has a tendency to repeat himself and therefore having the machine providing alternative solutions can be very useful. Unfortunately, I don't have that machine yet, but I would like to have it — a machine that basically criticizes what I'm doing and gives me suggestions. **It was like having someone else with whom you are composing there**. I think it is very interesting when two people are creating together, criticize each other and find a path that appeals to both. It is always different from a person who is alone, who tends to funnel more.

So I think having the computer with some autonomy to criticize I find that very useful. I also admit that having the computer autonomously composing is an interesting challenge from a scientific point of view and it can also be fun. From a scientific point of view it is a fantastic point of view.

**Artur -** I am not prejudiced against anything. I really like to experiment and saw myself using completely an automatic process. If the coordinates were designed by me and if the computer proposed a series of variations, I thought it was great, perfect. Actually, not perfect... but it is a possibility not only to produce visual graphic materials but also to automate my work in a more pragmatic way - to implement things for me so that I don't have to waste time.

# Q7. Have you been inspired by other designers 'work who used methods and tools different from yours? If so, do you think they helped you in any way to expand your own creativity?

```
Color code:
Has already been inspired by works
Reason / advantages of being inspired
```

**Ana -** Yes, without a doubt, in fact, I think it is always good to review what is already done, to see what others have already explored, how they thought and to get out of our thinking box a little. It is always good to look at a problem from a different point of view.

**Miguel -** I think anyone does that. At the academy, a design student ends up following this process. He learns about existing techniques, who uses them... and realizes what inspires him and leads him to try those same techniques / approaches. The learning process is to see how they do it or to try to discover other ways of doing it.

**Mariana -** Yes, without a doubt. Even during the bachelor, in the development of projects for some disciplines in which we researched the work of other designers and professionals from other areas. Try to understand what they created and how they created, understand what inspires them and what their work processes are. I think that inspires a lot.

**Catarina -** Yes, that's why I talked about research on the internet as a form of inspiration.

**Hugo -** Of course, this is very common, for example through sharing obtained at conferences or reading articles.

**João -** Yes, I think it always happens a little bit. In our profession, with our technical components, we can solve a design problem with any tool, but never in the same way. Although it is not completely canonical, there are tools that dictate a certain type of solution. This is inevitable, but we must make an effort to avoid it. On the other hand, without knowing certain types of tools, it will be difficult to come up with a certain type of solution.

**António -** Yes. It's like I said. We need to look at what already exists, combine and generate new solutions. In my case, I follow this same process. I combine tools and models to generate new solutions that, individually, these same tools and models cannot solve.

**Tiago -** Yes, I have already been inspired by many other designers who also develop these types of systems and creativity also goes to that level. Like any inspiration process, seeing things broadens your way of looking at a problem. When I see other approaches, I will start experimenting with other ways to solve the problem, for me this is creativity.

**Sérgio -** Yes, of course, everyone does. Even unconsciously when we are applying concepts and principles defined by other authors in the past. Of course they widened. I think that creativity is just that, it is the context where you are inserted, it is your knowledge, the books you read, your experiences. For example, a German has a completely different process / creativity from me.

**Nuno -** [see in other Qs]

**Amílcar -** What I normally also do is something I already do on a daily basis but now in a more directed way: it is to listen to a lot of music (see the state of the art). I collect inspirations about environments, certain forms of melody...

**Artur -** Our work, although we want to create breaks and explore new things, is always done taking into account the history of our discipline, the history of design. We respect the history of design.

In architecture there is a term that architects use a lot that is "architectural culture" (eg Souto Moura refers a lot to Mies Van Der Rohe). The work they do is done forward, but with a critical eye on the work of someone who has done things that they have liked in the past. This also happens to us, the exercise that we try to do is to look at works from the past and not at contemporaries, so that we don't have that temptation to do something and look like someone of our generation.

***[our team] Do you think this inspiration helps to expand the limits of your creativity?***

I mean, the ideal scenario is to look at all areas with a critical eye. We don't just look at graphic design. We look at art, contemporary art, modern art, conceptual art... and try to understand why artists did this and that. I think that to be a contemporary creator is to be attentive to history, to the various other creators, even to engineers, to writers, to the way of thinking, to philosophy, even to nature. We are looking at everything and being able to confront everything in a way. This seems very easy, but I think the important part of being inspired and influenced is trying to take a critical look.

## Q8. Co-creative systems are systems that involve computer programs that collaborate with human users in creative tasks. What do you think about systems of human-machine collaboration, co-creative systems, and the attribution of different degrees of autonomy / decision in the production of artefacts?

```
Color code:
Advantages of co-creative systems
I'd rather do it myself (compare with the advantages)
Depends on goals
Level of autonomy
Art vs design
```

**Ana -** It always depends on what we want to achieve. The introduction of the human being can be important to influence the creative process of the machine. It is also important that the machine has some autonomy to be able to surprise the human and to be able to generate things that could be necessary, but that the human being, perhaps, might not arrive. The weight of one and the other always depends on the desired result.

**Miguel -** I see advantages in this collaboration. It's the same as being working alone or talking to someone else. When talking to another person you just have an exchange of ideas that help you. Therefore, co-creative systems simulate this. **They are a partner that helps you to try to evolve an idea and that contributes in the same way or more than you**, when you simply just guide the evolutionary process of the system according to your preferences. Or you can draw and then the system draws and you draw again. Hence, the system's autonomy obviously affects the final result and the way the system works. I think that everyone is advantageous. The associated level of creativity is obviously dependent on the system's level of autonomy.

**Mariana -** I think it can have many advantages because of the unexpected factor, giving more or less power to the human or the machine, depending on what is intended.

**Catarina -** It depends on the system. **I see these systems as a partnership.** And **if, in fact, the system follows my goals**, everything is fine for me. On the contrary, if they generate something very different from what I intend, they are for research and not for creating a final product.

**Hugo -** Nothing against it. If machines exist and have certain capabilities that we do not have, we must take advantage of them. On the other hand, if there are other capabilities that we have that machines don't yet have, it is a way to get the best of both worlds.

**João -** In the strict design field, I think that solutions less constrained by the designer's will have more difficulties in solving problems, in the sense of design as a solution to a particular communication or space problem. It is clear that progress in this direction will be inevitable, especially with the tools of Artificial Intelligence. Now if we are talking about generative languages, making use of their randomness, the problem of artistic creativity is better than creativity in design. Here we are talking about the point of no use of the solution, which is something that is not very compatible with solving a design problem. If art is desirably useless, the design is not.

**António -** Yes, it can be interesting. If it is to use an autonomous system, I think that the idea of the user being able to manage the degree of autonomy and being able to influence the process is perhaps the best way to use these types of tools.

**Tiago -** I think that fully autonomous systems, in a few cases, can be successful. On the contrary, I believe that a co-creative system can create better solutions because they were worked on by the system and by the human co-author. At an early stage, the system can suggest solutions because it has more possibilities for random exploration and is not biased. The fact that it has a very wide spectrum can be an advantage. The human being ends up being contaminated by these solutions, he can remember other ideas or even just use one or more ideas that the computer suggested. Then he can start generating new ones, based on the ones he liked the most and, at some point, we can come up with more innovative, unique and creative solutions (whatever that may be).

About the different levels, we can have two extremes: little autonomy and we enter a production tool or something totally autonomous (ex: the initial version of Evotype), where the user can do little or nothing, could not express his preference. In terms of creativity, neither can help much. Using a fully autonomous system, it may be possible to generate an interesting solution, but in terms of creativity, it may be more interesting if the user is able to configure some aspects that he wants to value (eg final versions of Evotype, the user can choose whether he wants letters that have more straight lines or more curves). In the latter case, I am expressing some of my preferences, but I am not forbidding the system to show me some types of solutions.

**Sérgio -** I think that co-creative systems are great. They allow you to explore many paths quickly. Regarding autonomy, it can depend a lot on the phase of work in which it is applied or even on the type of work. The more creative the decision, at this point, the less autonomous the system has to be and vice versa.

**Nuno -** I think this has to do with the project. I think the project is going to say if it is a completely random thing, in which I enter some data and do not control the final result, or if I want to have control. I think it is the designer, according to the project he is doing, that defines the degree of autonomy of the machine in relation to the person.

**Amílcar -** Unfortunately I don't have that machine yet, but I would like to have it, a machine that critically criticizes what I'm doing and gives me suggestions. **It was like having someone else with whom you are composing there.** I think it is very interesting when two people are creating together, criticize each other and find a path that appeals to both. It is always different from a person who is alone, who tends to funnel more. (answer to another)

**Artur -** It has been shown that through artificial intelligence, through the sum of experiences... The machine manages to overcome Man on many issues. It can even overcome Man in medicine.

(I'm reading a book by Harani - "21 Lessons for the 21st Century". He talks about artificial intelligence exactly. He says that a good doctor is nothing more than an intelligent man who has been improving skills from personal experience, in contact with different situations, analyzing different patients.)

A computer well packed with artificial intelligence can do his job better. The computer can do better and faster than a doctor can make diagnostics (eg, analyze CT scans, tension...) and much more. I think that the computer does not surpass creativity, nor flexibility, nor the spirit of an artist and a more creative person.

**[our team] What do you think of the collaboration of co-creative systems? The user and the machine cooperate to develop an artifact.**

I think it is interesting. But I think that nowadays, to be programming such a thing, I would lose more time. It's quicker doing it myself than programming it to do it for me. Because the situations of creative work are always very different from each other. There is no pattern, at least from my experience. I make books, videos, websites for an architect, an artist, a curator, who have different ideas that I have to react to. Sometimes, I need to have an idea from one day to the next, so I think this is great, but for a type of work that is different from what I do. Now, a positive version: if I have a three-year project - book cover or record cover - this help will be very interesting and could have enormous potential (ex: if we can get references to the history of design, not just visualizations that detach themselves from the history of design).

# Q9. Fully autonomous systems are systems designed to generate creative artifacts without human assistance. Technological tools including machine learning and evolutionary approaches are used.
# Regarding autonomous systems, what do you think about letting a system like this design a book, a poster or a webpage with no human intervention? Do you think these types of systems could be useful in creative processes? Why?

```
Color code:
Support autonomous systems
Advantages of these systems
Art vs design
it depends / the problem is in the evaluation of the solution (related to the one above)
I don't know if I support / I don't know if they are capable (related to the two above)
```

**Ana -** There is always human intervention, even if it is the creator of the algorithm that creates the system. In addition, I think you should always have a human opinion, because the use may fall out of context. Ultimately, these systems are created for humans and therefore the evaluator will always be a human. Not that he has to intervene in the middle of the process, but there must be certain "checkpoints" for human evaluation or the user must have to validate at the end.

**Miguel -** I think so. Any type of technique is useful, depending on the purpose for which it is used. Everything is useful. It depends on the problem, the objective... I find it easy to find use in anything, looking in the right way. Yes, in autonomous systems, I see a lot of usefulness in using these.

**Mariana -** What is different about the machine is that it is not a human. And therefore, its way of thinking or creating... its reasoning… will automatically be different from yours and therefore it will generate something different. And I think that this can also be a form of inspiration. [part of the answer regarding the next Q was replaced accordingly]

**Catarina -** I think so. If it really is an autonomous system that works well and that manages to generate something coherent for its purpose... if it makes sense… then yes.

**Hugo -** Yes, again referring to the system I developed: We did a survey of several people (Coimbra and USA) to find out what they thought the benefits of the system were and how it could expand creativity. Some people said that these systems could help to "push" the so-called Blank Page Syndrome, thus starting the creative process. Others said that the system could generate something strange but that due to precisely these more unusual results, it led them to get ideas and follow paths that they would never think otherwise.

**João** - They can, in the sense of creativity in the artistic area, where it is guided by a conceptual approach and not problem-solving. But depending on the cases, yes. For example, in Web Design, the typification and optimization of some solutions are more than tested by humans. Through the use of these tools we can generate new knowledge and solutions. I accept that. In a short time, an innovative solution can be built in this way, using an algorithm. There are not many ways to design a chair (Siza Vieira), which therefore does not prevent a machine from not designing a good chair if we give it the proper knowledge about chair design.

**António -** Yes, I think it is interesting. Creating something entirely by the machine can generate interest, but perhaps in the process and not in the result. Yes, it could. It's like I already said. It could help to create initial shapes that you can combine.

**Tiago -** Yes and no. In the current state of the art, it may not be correct to place all systems in the same group because autonomous systems are further ahead in some domains than in others. For example, in terms of music, a more autonomous system may even make sense, we can get interesting results, we are at a more artistic level. On websites, although we may also have an art component, we have an even greater component of design, functionality and accessibility… so, or the system knows all the factors that are important (ex: target audience, what are the target audience's capabilities) and take this into account, or the outputs of the system will only be totally experimental and artistic artifacts, where the function is not valued. It always depends on the objectives, if the objective is to generate an experimental website, that's fine, but if we want to make a website on finance, I don't know to what extent an autonomous system (taking into account the current state of the art) has value. You may not be able to generate a function that has value, because in that context, value translates to functionality and not to how interesting and funny it was to click on the website.

A software can optimize much more than a human being, but the problem is not in optimizing, it is how you evaluate the solution. For example, what is it for you to be functional? IIf it is the number of clicks, I believe that a completely autonomous system would be able to design the best finance website. The problem is that the best finance site from a taxpayer's perspective is not just related to the number of clicks, but how the information is organized (what does it mean to be organized?) And how clear the information is (what does it mean to be clear?). The difficult part is to answer these questions, but if we can specify the quality of a solution, the system will be able to optimize it because it is mathematics. It always depends on the problem, the domain, and what is a good solution in the context.

If a good poster is one that uses less ink, an autonomous system can make the best poster. If a good poster is one that communicates information clearly and has a great emotional weight, I doubt that the autonomous system will be able to generate the best poster. It depends on how computationally we can evaluate the quality of the solution.

**Sérgio -** I think these systems are going to be the next creative leap in our area. Mainly to do routine tasks in many professional areas: photography, architecture, design...

**Nuno -** I don't have an immediate answer but I think so. Machines and technological solutions already do this and I already introduce this in my work process... already certain software solutions that help you automate certain issues (ex: if I need to resize a number of many images, I won't do it one by one, I program it). I see a middle ground there. I do not say that it is impossible, but I think that we are still far from a time when we input a number of images and texts and the computer pages a whole book, because I think there are issues of sensitivity that I think a machine will never learn, because a machine is not sensitive. I think it will take some time to the point where we have a complex design artifact fully developed by a machine.

*[our team] Ultimately, would you think it would be an asset to your creative process?*

But if there was a machine that did this automatically, where is the creative process at? I would no longer need to think.

*[our team] Maybe you can use such systems for generating ideas or generating content?*

Before, it took us a long time to reflect and then we were left with a single idea. Nowadays, I can try 20 different things in the space of an hour and, suddenly, that can already tell me which is the best way to go. So, my answer to that question is yes, because it speeds up the realization of the best idea. Because I see a series of possible variations and I can choose which one is the one that I think is most appropriate.

**Amílcar -** Yes, at least to see alternatives, for example. Let's imagine that somehow I managed to transmit some specifications to the machine and that it presented me with possible solutions, I could be inspired. Although what I liked most was having a machine that I was composing with. It was not entirely autonomous, it was a co-creative system. That was what I really liked.

**Artur -** when I think about it in society, I think it is a shot in the foot because a number of people would be unemployed. I am enriching the creator of this and then I am giving power to those who created it and the rest of the people are all poor. It is a pyramid that is neoliberalism in its state. Adobe is at the top of the pyramid right now in this area [we may have gotten off topic slightly]

There are a number of poor designers who are "putting their heads out" to pay Adobe and those guys fill their pockets with money. On top of that, I don't understand how Adobe, even with all the years of existence, has so many deficiencies (how do Illustrator and Adobe InDesign have different shortcuts?). I look at that for a while: "These guys are making fun of me for sure". For this reason, I do not advocate full automation and I think that those looking for a design studio do look for an interlocutor who reacts to a conversation, not a button to load.

*[our team] Maybe you can use these systems to create matters?*

Yes, if it matters just like I do. Sometimes I take lines, cross them and make photocopies. Then, I select and cut and paste. If it's to do this while I'm sleeping, if I don't have to make photocopies I think it's great. It is the same thing as asking a chef (who is going to get the lettuce from that garden with land made of a biological compost... going to get the carrot there... going to get the fish I don't know where and making it in a wood oven) if he wanted a machine to propose some frozen potato chips. So I think that computing, automation, is to help us with many components, but never to replace us [answer to the next Q].

## Q10. Do you think that the massification of such systems could threaten human creativity? Or instead, should we use them to evolve our capacities and, consequently, extend existing creative trends?

```
Color code:
Support autonomous systems
```

<span style="color:red">They Are not going to replace creativity</span>
<span style="color:orange">Can overcome humans / fashions</span>

**Ana -** <span style="color:green">I think that these systems can help human creativity</span>. I see this as <span style="color:red">a collaboration and not as a replacement.</span>

**Miguel -** First, I think the definition of computational creativity is a bit dangerous. It is not transparent. <span style="color:red">It is not easy to say what it is or to assess whether it is creative or not. So I think it is very difficult to reach a point where a system is more creative than a person.</span> I see the potential not in autonomous creative systems, but in the possibility of a user collaborating with a creative system, ultimately increasing the creativity of both parties. The sum of their creativity gives much more than each alone. <span style="color:red">A creative system alone is nothing. One person is something already.</span> I think that the potential of using a co-creative system ends up increasing creativity.

**Mariana -** <span style="color:green">Yes. We can only gain from this</span>. Sometimes the thought of the machine being able to replace the human scares me. In this case, replacing creativity. <span style="color:red">I believe not, because the machine's way of thinking. If we assume that the machine thinks, it will always be different from yours.</span> What is different about the machine is that it is not a human. And therefore, <span style="color:red">her way of thinking or of creating reasoning will automatically be different from yours and therefore it will generate something different.</span> And I think that this <span style="color:green">can also be a form of inspiration</span>. [same as above]

**Catarina -** <span style="color:red">I think it is always a finite system.</span> <span style="color:green">It is good to use</span>, <span style="color:red">but it will never replace us</span>, much less take away our creativity. <span style="color:red">It is a machine, it has no feelings or emotions and sometimes at work we need more of this human side.</span> I could do it myself and also use a system to generate results and do both in parallel and in the end I could compare the results, as if it was a partnership. However, I would never replace human work unless I had a lot of confidence about what came out of there and that it was great. But that way they didn't need me either. <span style="color:red">It would be more like a suggestion</span>. [taken from another question]

**Hugo -** I think you can find arguments for both points of view. It depends on the user's predisposition. On the other hand, <span style="color:green">it can precisely assist in the initiation of a process, allowing to stimulate the user's creative process and learn from it.</span> (He developed a system that generated riddles with a certain joke. One argument he had to defend the system was: this is not spectacular but maybe some of these jokes are generated by humans. Maybe these humans will lose their place, maybe because they are being a bit basic. In other words, <span style="color:orange">the system can produce artifacts that can surpass the human, whatever that is. In the same way that another system can combine several songs, as other humans may have already done in the same or in a better / worse way.</span>)

**João -** <span style="color:green">I think that the technologies that solve the most common problems and that let free our creativity for the most complex problems are always useful</span>. The fact that many common problems do not require the effort of designers, means that **designers are freer to solve and ask new questions**. <span style="color:red">Design is no longer just a discipline that solves problems but it is also a discipline that questions itself... that formulates problems.</span> This is because solving problems is getting easier and hence we are freer to make the design more complex... questioning it. <span style="color:red">Hence, designers are sought to help formulate problems and not just to solve them.</span>

**António -** I don't think it will threaten us. I don't think they will ever completely replace humans. At least in the evaluation process, they will not be able to surpass humans. In the creation process they are an asset.

**Tiago -** They will not threaten us, the designers themselves will have to adapt. We will talk about designers who do not create a unique solution, but rather systems. And the question can be translated into: "Will the designer create a system that will replace him himself?" I would say no, autonomous or co-autonomous systems are developed by the designer himself to generate solutions for him, which he himself will sell. He doesn't just have to sell the poster, he can charge for the time he spent creating the system. The risk can only exist because of trends. If people start to like generative posters a lot, non-programmer designers may feel a threat. However, it is all cyclical and sooner or later another one comes and no one likes generative posters anymore.

**Sérgio -** I don't think they are going to threaten our creativity and that we really should use them to expand our capabilities.

**Nuno -** I think it does not threaten human creativity, because we are just automating certain processes. There is talk on artificial intelligence and computational creativity. For me, the expression computational creativity generates discussion and I think this is natural because, for me, the computer has no creativity. The computer simulates creativity, which is quite a different thing. It is like artificial intelligence - the computer is not intelligent, the computer simulates certain functions that a human could do.

The human discovers things, often by accident, because he made a mistake and, suddenly, a door opens for another type of solution. A machine does everything it is supposed to. I think the beauty of creativity is in the unexpected, in knowing how to make use of it. It is knowing how to take advantage of the error and what happens unexpectedly, as part of the process. I don't put human beings ahead of creativity because, for me, human creativity is a pleonasm. Therefore, creativity is something intrinsic to humans and how humans make mistakes, get distracted or make decisions and have the sensitivity to look at the detail. They have an abstract thought that, probably, the machine will not be able to imitate - learn the error and internalize the error as a work process.
For the machine to be creative like humans, it has to learn from its mistakes… learn to make a mistake. Error is a very important thing in creative processes. In fact, so many times that I failed in the process and, many times, I failed with the final project, and I think about it a lot. I made many mistakes and I learned from these mistakes. The important thing is to know how to incorporate this later, in the next design project. The machines are not yet ready to do so.

**Amílcar -** We already have lots of autonomous systems producing music, these are the radios, the all "pimba" [Portuguese popular music], these are autonomous agents and these produce lots of crap that is not what prevents people from being creative. Then there is the social side... the enjoyment it gives to be playing together with others or even when we are listening and watching something, these are things that will not be lost. To me, for many autonomous agents out there, the human being does not seem to see his creation conditioned.

**Artur -** I think that everything increases our creative capacity. The fact that there is something that produces visual material for us, is increasing our critical spirit, so it is important. But now, if these tools are going to replace us? I don't believe it. And I will defend it until the end so these do to replace us because it is what I have been defending against all these years.

# B — Analysis

## Q1. What is your work-process (design-process) from gathering requirements to final-arts?

(i) FORMULATION - identify the problem, objectives to be achieved, requirements to develop the methodology, create a concept (12/12)
```
6 PhD students and all Professors (11/12)
Notes: One of the PhD students did not answer this directly but it was implied in other questions
```

(ii) RESEARCH - Investigating similar projects, searching the state of the art, relating and combining concepts (12/12)
```
5 PhD students, 1 Design Professor and 2 Informatics Professors (8/12)
Notes: All respondents said they did so in question 9
```

(iii) IMPLEMENTATION - Start developing, making various experiments, to use trial and error and iterative processes (11/12)
```
7 PhD students, 2 Design Professors and 2 Informatics Professors(11/12)
Notes: 1 Design Professor does not address the subject
```

(iv) Making documentation (1/12)
```
1 PhD student (1/12)
Notes: Only a PhD student talks about this issue, but most of the interviewees are researchers and therefore,
even if they do not consider the creative process, this part is done.
```

(v) Depends on the project (3/12)
```
1 PhD student and 2 Design Professors (3/12)
```

Conclusions:
(i) to identify the problem and objectives to achieve, to develop the methodology, to create a concept (12/12)
(ii) To investigate similar projects and the state of the art (12/12)
(iii) To start developing, making various experiments, to use trial and error and iterative processes (11/12)

**Some more unique topics**

- Limitations and detail in the work program are just as useful to satisfy these limitations as to work around them.
- One of the interesting things about the design process is the questioning of the work program.
- To what extent does the designer, as a result of the accumulation of knowledge he has, allow himself to question the program? The questioning of the program is accepted as one of the fundamental processes of the design process.
- *"Problem-solving is dictated by the problems themselves. There is no ideal solution to solve a design problem. The problem itself and the way the problem is interpreted dictates the solution. Different designers solve the same problem equally well or equally badly using different solutions. "*
João

**Conclusions**

They all have more or less the same work process, the order sometimes differs. There is a transversal process of identification of the problem, research of concepts, materials, solutions and then their combination leading to the realization of the solution. Very affected by the way the problem is interpreted and the knowledge acquired by the designer in his past and also during research.

# Q2. Which phases of your work-process imply creativity?

(i) Phases that imply creativity, from:
FORMULATION - identify the problem and objectives to achieve, requirements to develop the methodology, create a concept (12/12)
All PhD students and Professors (12/12)
Notes: A PhD student said that has the stage of requirements is less creativity

SEARCH - Investigating similar projects, searching the state of the art, relating and combining concepts (11/12)
7 PhD students, 2 Design Professors and 2 Informatics Professors (11/12)
Notes:
2 PhD students said that for a simple search, creativity is needed, not only in the way you are looking for, but also where and in what domains.
1 Design Professor does not refer to it.

IMPLEMENTATION - how the problem is approached and solved (11/12)
7 PhD students, 2 Design Professors and 2 Informatics Professors
Notes:
2 PhD students said that more in the implementation phase
1 Design Professor: does not refer

Notes:
António: All phases except the evaluation
Amílcar: All stages have a creative side if we want
João: Only refers to the formulation phase
Miguel: Creativity can arise anywhere. Even outside the process, creativity can arise in countless ways, actions and places.

(ii) The designer's background affects his creativity (3/12 of all interviews; ⅔ of senior designers)
The idea of creativity as a result of work, through previous practices, accumulates knowledge that influences future practices. The one who knows the subject best is most creative.
- *"Creativity is not a gift but the result of work. Technique, general knowledge and the way you work overlap with creativity, or better, it expands creativity."* Sérgio
- *"Sometimes I think it is not so much a matter of creativity in itself, I believe that if you are as or more creative the better you know the subject."* João
1 PhD student and 2 Design Professors

The support we choose can often be dictated by the novelty of new means of dissemination, and may often not be the most appropriate, which also leads to being creative in choosing the best dissemination approach João

**Conclusions**

The 3 main phases that involve creativity are Research, Formulation and Implementation.
The designer's background affects his creativity. Through past practices, knowledge accumulates that influences future practices. The one who knows the subject best is most creative. The way you interpret and approach the problem are also important aspects that have an effect on the application of your creativity in these 3 phases.

Implementation is the phase with the most creative focus.

For a simple search, creativity is needed, not only in the way you look for it, but also where and in what domains. (2/12)

Even outside the process, creativity can arise in countless ways, actions and places. (1/12)

# Q3. Which tools do you use in those phases? Do you usually use computational tools / systems during your creative process? If so, which of these tools / systems best support your creative process? Why?

(i) I use computational tools (12/12)
Information gathering / research tools (4/12)
`3 PhD students and 1 Design Professor`

    Organization / management tools (2/12)
    `1 PhD student ahd 1 Informatics Professor`

    Graphic design tools (5/12) - creativity support tools
    `3 PhD students and 1 Design Professor (all Adobe)`
    `Notes: 1 Design Professor speaks about other issues he uses`

    Creation / development tools (11/12) co-creative or standalone
    `7 PhD students, 3 Design Professors and 1 Informatics Professor`
        I develop my own tools `Miguel, ex: thesis`
        Java and Processing `Catarina, Ana, Mariana, António, Sérgio`
        JavaScript `Catarina, Sérgio`
        Python `Sérgio`
        Arduino `Ana`
        3D drawing tools - CAD, 3D renders `João`
        Sound tools - MaxMSP and others `Ana, Mariana, Amílcar`
        Code editors `António, Tiago`
        Algorithmic / unspecified tools? `Amílcar, João`
        `Notes:`
        `Artur, Nuno: talk about the use of the computer for creation (account still)`

    Version control tools
    `Tiago`
        Git

    Writing tools
    `Tiago, Hugo`
        Latex

    Reason for using the tools:
        Reason for using Adobe tools:
            - I rarely use alternative tools to keep from being integrated into the market and to stop having correspondence with other professionals, products. There is a certain hegemony in Adobe Studio that dictates this issue a lot. (`João`, on using Adobe tools)
            - It allows you to go back and understand how evolution was. Make editorial assembly easier `Artur`

Reason for using audio creation tools:
- It helps to search for existing patterns of a certain style, to join and then adapt. Adjustments, during the process, are only possible thanks to the tool. `Amílcar`

(ii) I use non-computational tools (6/12)
`3 PhD students, 2 Design Professors and 1 Informatics Professor`

Books, Archives `Catarina, Nuno`
Pencil, pen, paper `Catarina, Nuno, Hugo, António`
Photographs (tool?) `Artur`
Not specified `Ana`
Street `Catarina`

```
Notes:
Artur: to be a contemporary creator is to be attentive to philosophy, to nature (we will include)
Sérgio: Speak in books
Amílcar: Guitar
+Sérgio, Amílcar (talk about the other questions)
```

## Conclusions

All respondents say they use computational tools in their creative process. From implementation tools, where they give great focus, to tools to support the organization (version control and calendars) and information collection. Some of the interviewees, although they do not use tools or computer systems, say they take advantage of them through others ("I work with the tools that those who work with me work with.") `João`

Part of the interviewees (5/12) claims to use creativity support tools in the implementation of their work process. In addition, half of the interviewees also affirm that they use non-computational tools, but throughout the interview we realized that, in the majority, the interviewees continue to use analog methods (ex: books like research, paper and pen to support quick ideas and nature as inspiration). Depending on the needs of the project, different tools are used.

This part will move on to the conclusion of other questions:
- Some of the interviewees refer that they use Adobe tools to keep being integrated in the market and to correspond with other professionals (`João` and `Artur` in another question).
They also mention that these systems allow us to go back and understand how the process evolved and facilitate the editorial assembly (`Artur, Catarina` also talks about this later)
"I rarely use alternative tools to keep from being integrated into the market and to stop having correspondence with other professionals, products. There is a certain hegemony in Adobe Studio that dictates this issue a lot "
`João`
- These tools are useful because they can search for existing patterns and then adapt. (About musical creation)
`Amílcar`

## Q4. From your working-experience, what is the impact of the insertion of new tools in the creative process? (If you have already partnered and felt the benefits of introducing these tools, even if not directly, you can talk about it)

(i) Expanding results - expanding results, reaching further in experimentation, reaching unthinkable ideas, or starting from our ideas reaching new ones, the possibility of generating animated artifacts that are transformed (8/12)
`5 PhD students, 1 Design Professor and 2 Informatics Professors`

(ii) Facilities, process automation, speed (5/12)
`2 PhD students, 2 Design Professors and 1 Informatics Professor`

(iii) All work is saved (we can go back)
`1 PhD student and 1 Design Professor`

(iv) Dependence on the tool - we are more dependent on the tool than in the past. In contrast, in the past the process was conditioned by the accessibility of the tools - Reduced accessibility limited the process to other more accessible hand tools (letters of tracing, photocopying, photography, illustration and we get used to assembling things by hand)
`1 PhD student, 1 Design Professor and 1 Informatics Professor`

(v) Content sharing
`1 PhD student`

(vi) Uniformization
`1 Informatics Professor`

(vii) Introduction of new variables to be taken into account in the creative process (Ex April Greiman -> pixel assumed the pixel as an aesthetic factor that was part of his language) `Artur`

It allows you to expand the results, streamline and go further in the experimentation, come up with ideas that you were not having at the time, or through ideas that you have at the time, come up with other new ideas that you would not remember.
`3 PhD students`

Facilitating and speeding up experimentation and preventing and / or correcting errors.
`1 PhD student and 1 Design Professor`

It didn't change the design in terms of quality, but as Wim Crouwel says, it added speed `João`
We are much freer to create because the time-consuming and difficult technical processes have been passed on to the machine. Computing brought enormous speed to the process `João` and a separation between the technical component and the aesthetic, creative and exploratory component `António`

On the other hand, they make the process dependent on them. If the tools fail, the process fails. They make the whole process more intense due to its quick and easy access `Hugo`
there is a greater demand for deadlines. due to the introduction of mobile phones and computers in everyday life. The notion of time has changed a lot and I think for the worse because, culturally, we are losing things like patience, thoughtfulness and we are valuing a lot the immediacy.

The development is marked not only by the appearance of the software, but also by what comes with the software, the community. (e.g. Processing). Knowledge sharing and a creative method based on the creations of third parties appear, as if an expansion of these `Tiago` (From here we remove sharing as the central advantage of new technologies)

The evolution of graphic design is intrinsically linked to tools. `Sérgio`
New design problems are created, new spaces to explore leading to the growth of the profession. The whole profession grows dependent on the tools. (ex: typography. Technology allowed to use "lost" elements, without any typographic evolution, but rather an evolution in the possibilities of its application)

It is the speed and mode of reflection and execution of ideas. Shorter reflection times and shorter execution times due to the speed and communication of the whole process between both parties (client / professional) `Nuno`

 Acceleration in the human creation process
`Amílcar`

Introduction of new variables to take into account in the creative process (Ex April Greiman-> pixel assumed the pixel as an aesthetic factor that was part of his language)
`Artur`

**Conclusions**

(i) Increased speed of creation and exploitation of content, giving rise to alternatives that would not otherwise have been thought of. New variables are introduced in the creative process. On the other side of the creative spectrum, they can also lead to a standardization of the content produced. (12/12)
`All PhD students and Professors`

(ii) Control over the entire process and sharing it. A new perspective of content sharing and collaboration in projects arises and all creation / evolution can be controlled (ctrl + z, git, saves copies) (5/12)
`3 PhD students, 1 Design Professor and 1 Informatics Professor`

(iii) Its ease of use and access, created a dependency on the creative process, as in the past, in contrast, its difficult accessibility dictated the limitation of the tools used for a given project (3/12)
`1 PhD student, 1 Design Professors and 1 Informatics Professor`

# Q5. Regarding the impact on creativity, how do you think that computational tools / systems differ from non-computational tools / systems, that is, manuals (pros and cons)?

**Digital Pros**

- Speed, time saving (12/12)
All the PhD students and Professors

- Increased freedom of creative exploration and control over your process (3/12)
1 PhD student, 1 Design Professor and 1 Informatics Professor.

- Ease of access and information management (2/12)
1 PhD student and 1 Informatics Professor

- Specialization of tools (Focused to solve set problems) (1/12)
1 PhD student

- Expansion of the study areas of several professional areas (1/12)
1 PhD student

**Cons Digital**

- Bleached capacity for reflection in favor of experimentation facilitated by the PC (2/12)
2 Design Professors

- Uniformization (1/12)
1 Informatics Professor

**Analog / Manual Pros**

- The whole process of execution and exploration of the domain is much more thoughtful and reflected a priori (3/12)
1 PhD student and 2 Design Professors

- Manipulation closer to the real world, with unique qualities (final artefact) (2/12)
2 PhD students

**Cons Analog / Manual**

- Time-consuming and difficult technical processes dictated much of the success of the work, not allowing great freedom / possibility for creative exploration (1/12)
1 Design Professor

- The use of Analog and / or Digital Systems / Tools depends strongly on the context, type of project (3/12)
3 PhD students

- The lack of habituation to a certain type of tools (analogue / digital) can lead to different solutions than those that usually use them (1/12)
  `1 PhD student`

**Conclusions**

The advantages of the tools (a / d) prove to be of added value depending on the context and type of project.
Speed and time-saving in the production of content provided its ease of access and control over the information process. There are also advantages in combining the two types of tools in order to generate more experimental and less standardized results.
The evolution of the tools led to the expansion of the study areas of the professional areas.

## Q6. Would you use a new co-creative or fully autonomous tool/system to help you with your creative process? What would lead you to use a new tool?

(i) Everyone would use tools of this kind. Because:

- Necessity and / or Curiosity (2/12)

`2 PhD students`

- The unexpected (3/12)

`2 PhD students and 1 Design Professor`

- Creative Inspiration / Suggestion (as complementary creative content) (7/12)

`4 PhD students, 1 Design Professors and 2 Informatics Professor`

- Expansion of features and solutions taken from current tools (1/12)

`1 Design Professor`

- Dynamism (1/12)

`1 Design Professor`

## Q7. Have you been inspired by other designers 'work who used methods and tools different from yours? If so, do you think they helped you in any way to expand your own creativity?

Have already been inspired (9/12)

`7 PhD students, 1 Design Professor and 1 Informatics Professor`

Search for inspiration in other areas as well (1/12)

`1 Design Professor`

(i) Advantages:

- Broaden horizons (1/12)

`1 PhD student`

- Know techniques (5/12)

`4 PhD students and 1 Informatics Professor`

- See what is done to make things different (2/12)

`2 PhD students (combining things)`

- Criticize what was done (more or less the same) (1/12)

`1 Design Professor`

### - Inspiration (4/12)
`2 PhD students, 1 Design Professor and 1 Informatics Professor`

### - Knowing Tools (1/12)
`1 Design Professor (We need to know certain types of tools to arrive at certain solutions)`

### There are tools that determine a certain type of solution (1/12)
`1 Design Professor`

`notes:` **creativity is just that, it is the context in which you are inserted, it is your knowledge, the books you read, your experiences.** `Sérgio`

## Q8. Co-creative systems are systems that involve computer programs that collaborate with human users in creative tasks. What do you think about systems of human-machine collaboration, co-creative systems, and the attribution of different degrees of autonomy / decision in the production of artefacts?

(i) Level of autonomy:

- Having enough to overcome the human and reach different results (2/12)

1 PhD student and 1 Informatics Professor

- Human guide the process can be important (7/12)

6 PhD students and 1 Design Professor

- More autonomy more creative (1/12)

1 PhD student

- Less autonomy more creative (2/12)

2 PhD students

- It depends on the goals (6/12)

5 PhD students and 1 Design Professor

(ii) In favor (4/12)

3 PhD students and 1 Informatics Professor

(iii) Utilities:

- Generate new ideas in partnership (7/12)

5 PhD students and 2 Informatics Professor

- Explore (4/12)

3 PhD students and 1 Design Professor

- Create (if guided by the user) (2/12)

2 PhD students

- Evaluate Human work, and suggest new solutions (1/12)

1 Informatics Professor

- Automate graphic image for projects of many years but not small (1/12)

1 Design Professor

- The computer will never surpass human creativity (1/12)

1 Design Professor

**Conclusions**

Most agree that the human guide the process is advantageous (more in favor of co-creatives) The desired levels of autonomy depending on the project. Systems can be used to discover new paths. Some people imagine themselves collaborating with autonomous systems (the system suggests options; the process turns out to be co-creative).

**Q9. Fully autonomous systems are systems designed to generate creative artifacts without human assistance. Technological tools including machine learning and evolutionary approaches are used.
Regarding autonomous systems, what do you think about letting a system like this design a book, a poster or a webpage with no human intervention? Do you think these types of systems could be useful in creative processes? Why?**

(i) Does not believe in fully autonomous systems (there is always a need for human intervention) (2/12)
```
1 PhD student and 1 Design Professor (we are far from that)
```

(ii) In favor of freelancers (6/12)
```
4 PhD students, 1 Design Professor and 1 Informatics Professor
```

    In the process yes, in the final product NO (co-creative?) (4/12)
```
1 PhD student, 2 Design Professors and 1 Informatics Professor
```

(iii) Utilities

    - Everything is useful for a particular problem (3/12)
```
2 PhD students and 1 Design Professor
```

    - Think differently than a human (see new solutions) (6/12)
```
2 PhD students, 2 Design Professors and 2 Informatics Professors

Notes: 1 PhD student, 2 Design Professors and 1 Informatics Professor believed that
"Even if the results are strange it can lead to other ideas"
```

    - Helping in the process (6/12)
```
3 PhD students, 1 Design Professor and 2 Informatics Professors
```

    - More in art than functional design (2/12)
```
1 PhD student and 1 Design Professor
```

    - Do routine tasks (1/12)
```
1 PhD student
```

(iv) The problem may be the assessment (3/12)
```
2 PhD students and 1 Design Professor
```

(v) The problem is not having sensitivity or feelings (2/12)
```
1 PhD student and 1 Design Professor
```

**Conclusions**

Respondents are in favor of using sist. autonomous to give suggestions in the creative process (collaboration) Useful for generating new ideas. Most useful when the evaluation of the resulting artifacts is quantitative and objective. Useful for automating / replacing routine tasks.

# 10. Do you think that the massification of these systems could "threaten human creativity"? Or, should we instead use them to evolve our capabilities and, consequently, extend existing creative currents?

(i) In favor of co-creative or autonomous systems (8/12)
`6 PhD students, 1 Design Professor and 1 Informatics Professor`

    - Co-Creatives (1/12)
`1 PhD student and 1 Design Professor`

    - Autonomous (0/12)

    - Both (1/12)
`1 PhD student`

    - Apparently both (it is not clear) (6/12)
`5 PhD students and 1 Informatics Professor`

(ii) Supports automation (systems that solve common problems and unleash creativity for more complex things, but it is unclear whether these systems include co-creative or autonomous systems.) (1/12)
`1 Design Professor`

(iii) Co-creative systems can increase creativity (7/12)
`5 PhD students, 1 Design Professor and 1 Informatics Professor`

(iv) systems (co-creative and autonomous) will not replace human creativity (11/12)
`7 PhD students, 3 Design Professors and 1 Informatics Professor`

(v) Fully autonomous systems can threaten human creativity (people may just want to press a button and the system will generate) (1/12)
`1 Informatics Professor`

(vi) Can substitute for simple tasks (1/12)
`1 Informatics Professor`

(vii) Designers have to adapt (2/12)
`1 PhD student and 1 Design Professor`

**Conclusions**

The majority thinks it will not be replaced. Both are more capable in certain tasks - let's complement each other. Some people think it is good for machines to take on some human tasks as humans can use their creative abilities to explore other paths or areas.
`(for example, if the machines do static graphic design, the designers will start to evolve more the part of the interactive or dynamic design) - nobody said, it is an example.`

# Axial Coding Model

**1 - Casual Conditions**
What influences the central phenomenon, events, incidences, happenings

**2 - Phenomenon**
The central idea, event, happening, incident about which a set of actions or interactions are directed at managing, handling or to which the set of actions is related.

**3 - Strategies**
For addressing the phenomenon. Purposeful, goal-oriented

**4 - Context**
Locations of events. Where, when, with whom.

**5 - Intervening Conditions**
Shape, facilitate or constrain the strategies that take place within a specific context

**6 - Consequences**
Outcomes or results of action or interaction, result from the strategies

|  |  | 4<br>-Type of project<br>-Environment<br>Curiosity(leasure)<br>-Necessity(professional)<br>-Professional Integration<br>-Extent creative space of solutions<br>-Organization |  |
|---|---|---|---|
| 1<br><br>- Use of new/different? type of systems/tools | 2<br>- Emergence of new possibilities<br>- Expand the creativity limits | 3<br>-Making use of co-creative and fully autonomous systems/tools<br>-Combine multiple systems/tools<br>-Share information<br>- Seek inspiration from both human and computational sources<br>estratégias, ferramentas | 6<br>- Speed and production increase<br>- Technological dependencies<br>-Increased project control and management<br>- New areas of study<br>-Decay of creative thinking |
|  |  | 5<br>-Time<br>-Knowledge Background<br>-Systems/Tools Hegemony<br>-Technological/Monetary Accessibility<br>-Partnerships<br>modus-operandi |  |